\documentclass[sigplan,10pt]{acmart}
\usepackage{booktabs}   
\usepackage{subcaption} 

\usepackage{multirow}
\usepackage{algorithm,algorithmicx,algpseudocode}
\usepackage{pifont}
\usepackage{color}
\usepackage{fontawesome}

\usepackage{textcomp}
\usepackage{xcolor}
\usepackage{booktabs}

\usepackage[]{hyperref}
\acmConference[PPoPP'25]{ACM SIGPLAN Conference on Programming Languages}{2025}{Las Vegas, NV, USA}
\acmYear{2025}
\acmISBN{} 
\acmDOI{} 
\startPage{1}

\setcopyright{none}

\begin{document}

\title[]{FlashSparse: Minimizing Computation Redundancy for Fast Sparse Matrix Multiplications on Tensor Cores}         

\author{Jinliang Shi}
\email{shijinliang@bupt.edu.cn}
\affiliation{%
  \institution{Beijing University of Posts and Telecommunications}
  \country{Beijing, China}
}

\author{Shigang Li}
\authornote{Corresponding author: Shigang Li.}
\email{shigangli.cs@gmail.com}
\affiliation{%
  \institution{Beijing University of Posts and Telecommunications}
  \country{Beijing, China}
}

\author{Youxuan Xu}
\email{youxuanxu@bupt.edu.cn}
\affiliation{%
  \institution{Beijing University of Posts and Telecommunications}
  \country{Beijing, China}
}

\author{Rongtian Fu}
\email{rongtianfu@bupt.edu.cn}
\affiliation{%
  \institution{Beijing University of Posts and Telecommunications}
  \country{Beijing, China}
}

\author{Xueying Wang}
\email{wangxueying@bupt.edu.cn}
\affiliation{%
  \institution{Beijing University of Posts and Telecommunications}
  \country{Beijing, China}
}

\author{Tong Wu}
\email{2023140627wt@bupt.edu.cn}
\affiliation{%
  \institution{Beijing University of Posts and Telecommunications}
  \country{Beijing, China}
}

\begin{abstract}
Sparse Matrix-matrix Multiplication (SpMM) and Sampled Dense-dense Matrix Multiplication (SDDMM) are important sparse operators in scientific computing and deep learning. Tensor Core Units (TCUs) enhance modern accelerators with superior computing power, which is promising to boost the performance of matrix operators to a higher level. However, due to the irregularity of unstructured sparse data, it is difficult to deliver practical speedups on TCUs. To this end, we propose FlashSparse, a novel approach to bridge the gap between sparse workloads and the TCU architecture. Specifically, FlashSparse minimizes the sparse granularity for SpMM and SDDMM on TCUs through a novel swap-and-transpose matrix multiplication strategy. Benefiting from the minimum sparse granularity, the computation redundancy is remarkably reduced while the computing power of TCUs is fully utilized. Besides, FlashSparse is equipped with a memory-efficient thread mapping strategy for coalesced data access and a sparse matrix storage format to save memory footprint. Extensive experimental results on H100 and RTX 4090 GPUs show that FlashSparse sets a new state-of-the-art for sparse matrix multiplications (geometric mean 5.5x speedup over DTC-SpMM and 3.22x speedup over RoDe).
\end{abstract}



\maketitle

\section{Introduction}
Sparse matrix-matrix multiplication (SpMM) and sampled dense-dense matrix multiplication (SDDMM) are two major sparse operators used in various fields, such as scientific computing~\cite{anzt2015accelerating,blei2003latent,lan2014sparse} and graph neural networks (GNNs)~\cite{scarselli2008graph,kumar2022influence,bui2022spatial,jiang2022graph,almasan2022deep,piao2022sparse}. 
For example, in GCN~\cite{kipf2016semi}, the feature aggregation of neighbor nodes (i.e., graph convolution) can be computed as an SpMM, while in AGNN~\cite{thekumparampil2018attention} and GAT~\cite{velivckovic2017graph}, the attention between graph nodes can be computed as an SDDMM. 
Since these sparse operators often lead to performance bottleneck, the acceleration of SpMM and SDDMM on GPUs has been widely studied. One class of research works~\cite{sputnik, RoDe, wang2021gnnadvisor, huang2020ge} focus on accelerating sparse operators on GPU CUDA cores.
Gale et al. proposed Sputnik~\cite{sputnik}, a 1-dimensional tiling scheme to hierarchically decompose and map the sparse computation to CUDA cores.
This tiling scheme greatly improves the data locality and occupancy.
RoDe~\cite{RoDe}, the current newest work on CUDA cores, aims to address the issue of load imbalance in Sputnik.
RoDe first splits sparse matrices into long rows and short rows. Long rows are further divided into finer-grained groups. This load-balancing method enhances concurrency, especially for extremely unevenly distributed sparse matrices.

On the other hand, the recently emerging Tensor Core Units (TCUs) account for a major part of the computing power (much higher than CUDA cores) on modern GPUs.
TCUs were first introduced in NVIDIA Volta GPUs to accelerate \textbf{m}atrix \textbf{m}ultiplication and \textbf{a}ccumulation (\textbf{MMA}) operations.
TCUs have been widely leveraged to speed up scientific computing and deep learning workloads, but mainly for dense matrix operations~\cite{li2022research,khurana2023natural} or structured sparse matrix operations~\cite{cusparselt}. For example, cuSPARSELt utilizes the natively-supported sparsity on TCUs~\cite{SparseTC} to achieve the double
peak performance compared to the dense counterparts. But it imposes strict constraints on the sparse pattern (i.e., 2:4
structured sparsity) with a sparsity ratio constrained to 50\%, which limits its usability. 

\begin{figure}[htbp]
  \centering
  \includegraphics[width=\columnwidth]{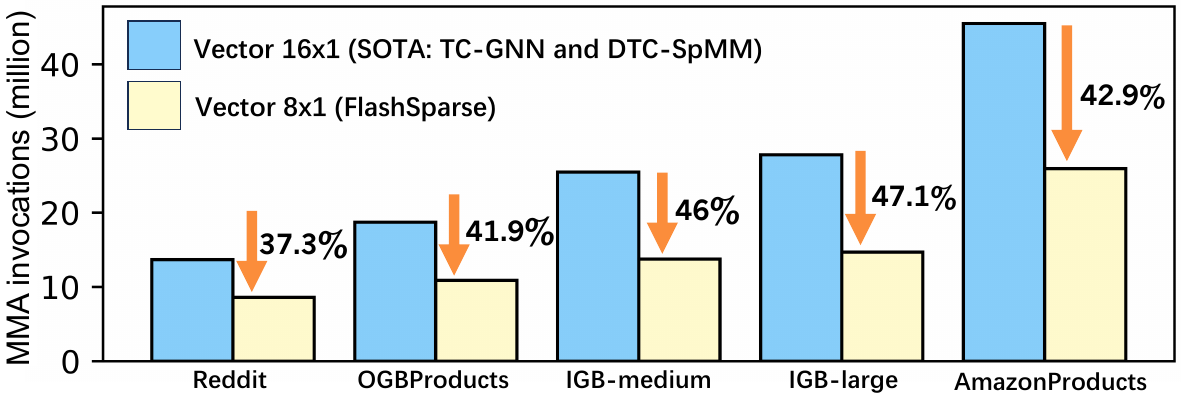}
  \caption{The number of MMA invocations under 16x1 and 8x1 nonzero vector sizes in SpMM. Note that the unit used for IGB-large is \textit{ten millions} for clear presentation.}
  \label{fig:motivation}
\end{figure}

However, sparse matrices in many real-world applications~\cite{sala2010brief,kipf2016semi,thekumparampil2018attention,velivckovic2017graph,SuiteSparse} exhibit high sparsity ratio (e.g., >99\%) and unstructured pattern, which is far different from the structured sparsity natively supported on TCUs. The major challenge in utilizing TCUs for accelerating these sparse operators lies in the mismatch between the sparse workloads and the TCU architecture. 
A naive implementation may identify nonzero blocks in sparse matrices at the granularity of operand shapes (e.g., 16$\times$8) supported by the MMA instruction, but this leads to severe resource waste since most of the values in these blocks are zeros. 
Several research efforts~\cite{wang2023tc,dtc,magicube,chen2021efficient} have gone into designing more efficient sparse matrix storage format and kernel optimizations to improve the resource utilization. For example, TC-GNN~\cite{wang2023tc} and the latest work DTC-SpMM~\cite{dtc} identify 16$\times$1 nonzero vectors in sparse matrices and then concatenate these vectors into the block shape supported by the instructions (MMA or WMMA~\cite{mma}) on TCUs. As a result, the all-zero vectors are eliminated from the computation, resulting in higher utilization. 
However, the design philosophy of TC-GNN and DTC-SpMM necessitates the nonzero vector size of 16$\times$1, which is still too large to be efficient, since a large portion of the values in the nonzero vectors are zeros and the computing power is wasted.

To this end, we propose a novel approach, FlashSparse, which bridges the gap between the sparse workloads and the TCU architecture. Through an exquisite algorithm design and highly-optimized implementation, FlashSparse can minimize the granularity of nonzero vectors to 8$\times$1. Note that the smaller vector size in FlashSparse is not achieved by sacrificing the computing power of TCUs, but from a sophisticated hardware-software co-design. In Figure~\ref{fig:motivation}, we compare the number of MMA invocations for an SpMM operator on sparse matrices generated from graph data, when using different vector sizes (16$\times$1 in TC-GNN and DTC-SpMM, 8$\times$1 in FlashSparse). 
The number of columns of the dense matrix is 16.
We observe that the 8$\times$1 vector size can reduce the number of MMA invocations by an average of 43\% compared to 16$\times$1, which can be directly translated into a remarkable reduction of computation and data access cost for sparse operators on TCUs. 
Our main contributions are:
\begin{itemize}
 \item We identify the key factor of performance limitation in state-of-the-art works using TCUs to accelerate sparse operators, namely the high redundancy of computation and data access caused by the large nonzero vector size imposed by the algorithm design.
 
 \item We propose FlashSparse, which minimizes the nonzero vector granularity to 8$\times$1 for SpMM and SDDMM on TCUs through a novel swap-and-transpose MMA computation strategy.
 \item While implementing the sparse kernels in 8$\times$1 vector size, FlashSparse adopts a memory-efficient thread mapping strategy for coalesced data access, bringing a significant reduction of memory transactions.
 
 \item Extensive experiments on H100 and RTX4090 GPUs show that FlashSparse sets a new state-of-the-art for both sparse kernels (e.g., geometric mean 5.5x and up to 25.26x speedup over DTC-SpMM on 515 different sparse matrices) and end-to-end GNN applications (e.g., geometric mean 1.79x and up to 2.83x speedup over the latest version of DGL~\cite{wang2019deep}). 
\end{itemize}

\section{Background and Motivation}
\subsection{Tensor Core Units}
Tensor Cores~\cite{tensor} are specialized computing units in modern GPUs to accelerate the operations of matrix multiplication and accumulation (MMA). 
Compared to CUDA cores, TCUs deliver superior computing power, but only for MMA. To program
on TCUs, CUDA provides two warp-level APIs for matrix multiplication and accumulation, including WMMA (C++ API) and MMA (low-level quasi-assembly). Note that in CUDA, threads are scheduled in warps and each warp has 32 threads.
As illustrated in Table~\ref{tab:shapes}, different operand shapes are supported for the two APIs.
WMMA-TF32 is used in TC-GNN, while MMA-TF32 is used in DTC-SpMM.
Compared to WMMA, MMA instructions enable finer-grained and more flexible matrix operations for sparse operators. 
TC-GNN uses WMMA with $m16n16k8$ for TF32 while DTC-SpMM employs  MMA with $m16n8k8$ for TF32. In FlashSparse, we utilize MMA with $m16n8k4$ for TF32 and MMA with $m16n8k8$ for FP16.

\setlength{\tabcolsep}{2pt}
\begin{table}[]
    \centering
    \caption{The shapes of WMMA and MMA on Tensor Cores}
    \begin{tabular}{ccc}
    \toprule
    Type & Precision & Operand shapes\\
    \midrule
    WMMA & TF32 &  m16n16k8  \\
    MMA & TF32 &  m16n8k4, m16n8k8\\
    MMA & FP16 &  m16n8k8, m16n8k16\\
    \bottomrule
    \end{tabular}
    \label{tab:shapes}
\end{table}



\begin{figure}[htbp]
  \centering
  \includegraphics[width=\columnwidth]{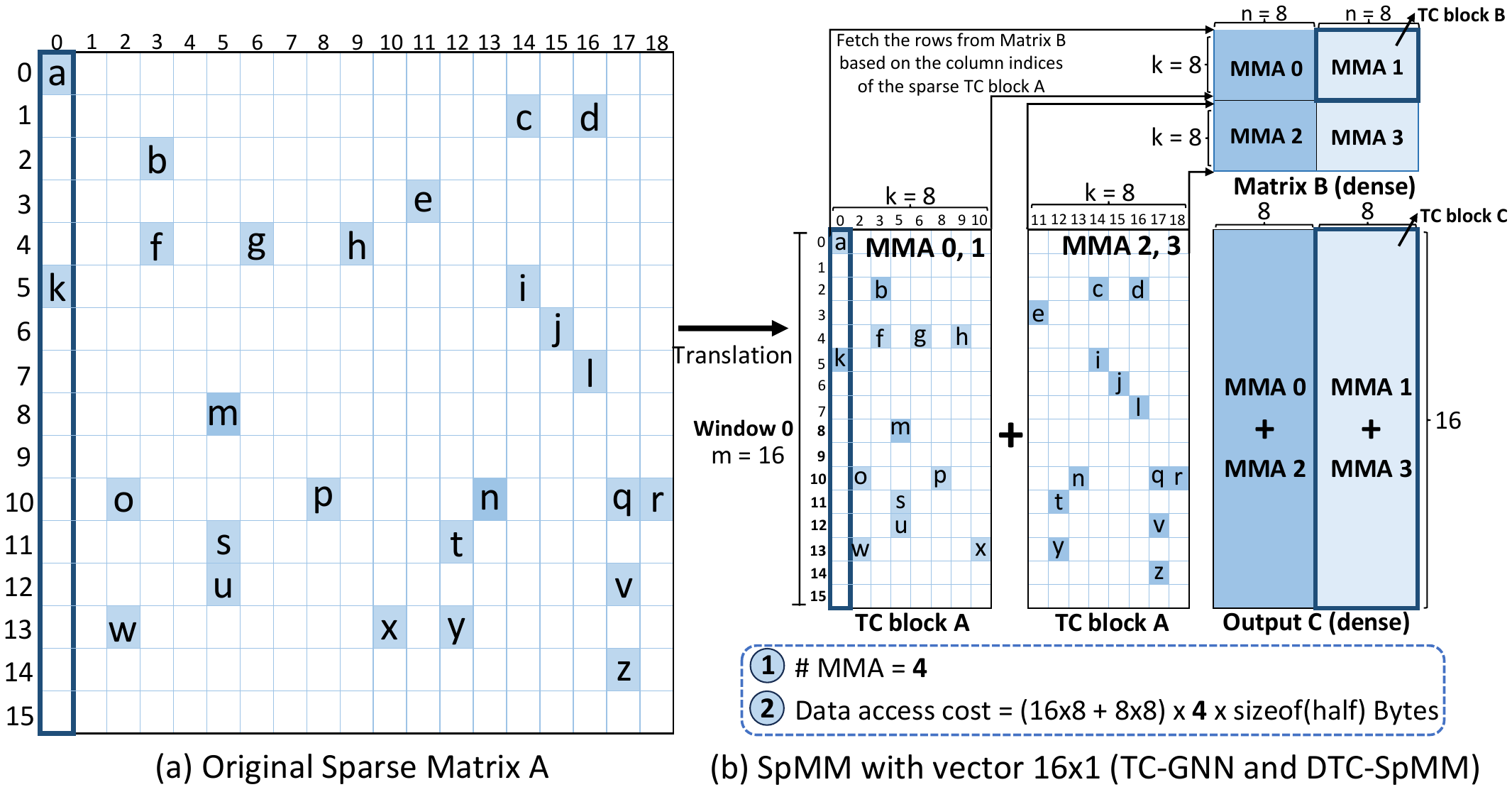}
    \caption{SpMM on TCUs with nonzero vector size of 16$\times$1. The operand shape of MMA is m16n8k8.}
    \label{fig:16times1}
\end{figure}

\subsection{Sparse operators on TCUs}
\label{sec:currentwork}
We take SpMM (a sparse matrix A multiplied by a dense matrix B) on TCUs as an example to describe the process of nonzero vector partition and MMA computation in current state-of-the-art works.
With FP16 precision, MMA requires the matrix shape to be $m16n8k8$.
As illustrated in Figure~\ref{fig:16times1} (a), the original sparse matrix A is partitioned into 16$\times$1 (i.e., 16 rows and 1 column) vectors according to the $m$ dimension (e.g., 16) of MMA, as in TC-GNN and DTC-SpMM. A row of nonzero vectors in the matrix is called a \textit{window}. Any vector that contains at least one nonzero element is called a \textit{nonzero vector}. Each $k$ (e.g., 8) nonzero vectors in a window are combined to form a 16$\times$8 TC block A, which serves as the left operand of MMA, as shown in Figure~\ref{fig:16times1}(b). Note that the blank space in the sparse TC blocks of matrix A are filled with zeros. 
Then, $k$ (e.g., 8) rows of matrix B with length $n$ (e.g., 8) are fetched based on the column indices of the nonzero vectors in TC blocks of A, which forms a 8$\times$8 TC block B serving as the right operand of MMA.
The intermediate results (such as the output of MMAs 0 and 2) are accumulated to the output block C with the size of $m \times n$ (e.g., 16$\times$8).


\subsection{The impact of nonzero vector size}
In the previous section, we present how existing works (e.g., TC-GNN and DTC-SpMM) adapt SpMM to TCUs. Particularly, their algorithm design method necessitates the nonzero vector length equal to the $m$ dimension of MMA (i.e., 16), as shown in Figure~\ref{fig:16times1} (b). However, many real-world applications exhibit high sparsity and irregularity, which causes most of the elements in the 16$\times$1 nonzero vectors to be equal to zero.
Table~\ref{tab:redundant} compares the number of zero values in the nonzero vectors at different vector sizes (16$\times$1 and 8$\times$1) using real-world datasets. 
We can observe that, when using 16$\times$1 vector size, the number of zero values in the nonzero vectors is much higher than the nonzero values, from 5.6x to 11.4x. Since these zero values have no contribution to the final results of MMA, most of the computing power of TCUs is wasted. In contrast, if we simply reduce the vector size to 8$\times$1 without considering the matrix shapes supported by MMA, the number of zero values in the nonzero vectors is remarkably reduced by approximately 50\% for all datasets in Table~\ref{tab:redundant}. However, naively using a smaller vector size without considering the computation pattern on TCUs makes no sense, since the mismatch between the block shape formed by nonzero vectors and the matrix shape supported by MMA directly causes severe computing power waste. Therefore, although a smaller vector size is promising to reduce the number of zero values, perfectly matching the smaller vector size to the TCU computation pattern is very challenging. In the following, we will present how FlashSparse achieves to efficiently minimize the size of nonzero vectors while fully utilizing the computing power of TCUs.


\setlength{\tabcolsep}{0.8pt}
\begin{table}[]
    \centering
    \caption{The number of zero elements in the nonzero vectors at different vector sizes (Unit: millions).}
    \begin{tabular}{ccccc}
    \toprule
    \multirow{2}[2]{*}{Graph datasets} & \multirow{2}[2]{*}{\#Nodes} & \multirow{2}[2]{*}{\#Edges} &\multicolumn{2}{c}{\#Zeros in nonzero vectors} \\
    \cmidrule(lr){4-5}
    & & & \textbf{16$\times$1} vector &  \textbf{8$\times$1} vector\\
    \midrule
Reddit & 0.23 & 114.8   & 762.6 & 435.7\\
OGBProducts & 2.5 & 126.1   & 1072.1 & 570.6\\
IGB-medium & 10 & 129.9   & 1502.1 & 751.2\\
IGB-large & 100 &1,323.5   & 16,468.8 & 8,085.4\\
AmazonProducts & 1.6 & 264.3   & 2648.8 & 1396.9\\
    \bottomrule
    \end{tabular}
    \label{tab:redundant}
    \vspace{-1em}
\end{table}

\section{FlashSparse}
\subsection{Overview}
\begin{figure}[htbp]
  \centering
  \includegraphics[width=\columnwidth]{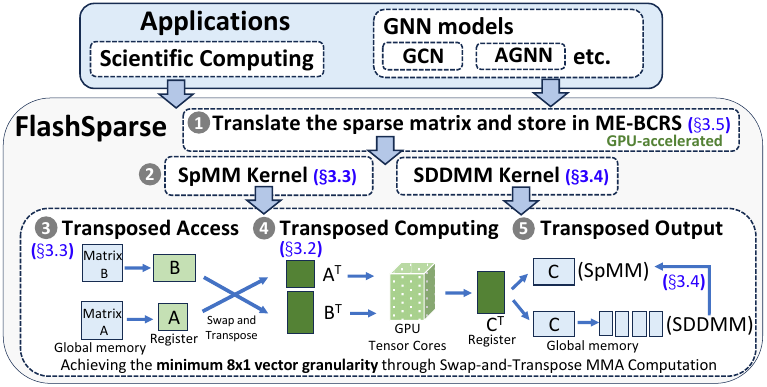}
  
  \caption{Overview of FlashSparse.}
  \vspace{-1em}
    \label{fig:overview}
\end{figure}

FlashSparse is an approach to accelerate SpMM and SDDMM on TCUs by achieving the minimum vector granularity.
As shown in Figure~\ref{fig:overview}, FlashSparse consists of several key components aimed at   maximizing the performance of sparse operators on TCUs.
The workflow of FlashSparse involves two main parts: sparse matrix translation and swap-and-transpose MMA computation.
Initially, the sparse matrix is translated into sparse TC blocks based on the vector size.
These sparse TC blocks are then converted into a memory-efficient storage format optimized for TCUs.
Note that the matrix translation process leverages CUDA for parallel processing on the GPU.
For the kernel implementation, the sparse operators (SpMM and SDDMM) adopt a swap-and-transpose MMA computation strategy to achieve the minimum vector size of 8$\times$1.
Specifically, the swap-and-transpose strategy involves the steps of operands swapping, transposed access, transposed computation, and transposed output.
Benefiting from the minimum 8$\times$1 vector granularity, FlashSparse significantly reduces both computation redundancy and data access costs for sparse operators using TCUs.

\subsection{Swap-and-transpose MMA computation}
\begin{figure}[htbp]
  \centering
  \includegraphics[width=\columnwidth]{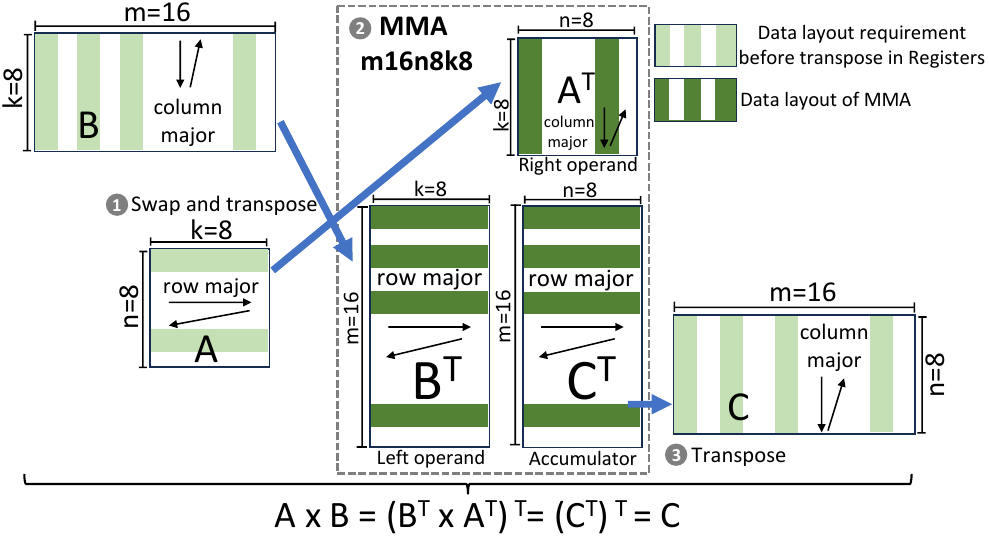}
  \caption{The swap-and-transpose MMA computation.}
  \label{fig:transpose}
\end{figure}

As we mentioned above, the vector granularity of existing works is equal to the $m$ dimension of the \textbf{left operand} in MMA.
However, the smallest $m$ is 16 as shown in Table~\ref{tab:shapes}, which is too large to be efficient. In contrast, the $n$ dimension of the \textbf{right operand} in MMA is 8 (half of $m$). Therefore, we propose to leverage the mathematics $A\times B$=C $\Rightarrow B^{T}\times A^{T}$=$C^{T}$ to execute the MMA in such a way that the two operands are exchanged and the vector size changes to be the smaller dimension $n$.
In Equation~\ref{equ:trans} and Figure~\ref{fig:transpose}, we give an overview of the swap-and-transpose MMA computation strategy:
\begin{equation}  \label{equ:trans}
\begin{gathered}
   A \times B =  \left(B^{T} \times A^{T}\right)^{T} = \left(C^{T}\right)^{T} = C
\end{gathered}
\end{equation}

where TC blocks A and B are the target data blocks to be accessed from global memory; TC block C is the result that needs to be written to global memory; $A^{T}$, $B^{T}$, and $C^{T}$ (the transposed TC blocks) are the actual operands of MMA in registers.
By leveraging the swap-and-transpose strategy, the sparse TC block A is transposed into $A^{T}$ to serve as the \textbf{right operand} ($k \times n$) of MMA, while the dense TC block B is transposed into $B^{T}$ to serve as the \textbf{left operand} ($m \times k$) of MMA.
Therefore, $n=8$ can be leveraged as the vector size to partition the sparse matrix without sacrificing the computing power of MMA.
However, during the swap-and-transpose MMA computation, the two input operands require positional swapping and transpose. 
This results in data layout mismatch between the requirement in registers and the input and output matrices in global memory.
Therefore, efficiently incorporating the swap-and-transpose MMA computation into SpMM and SDDMM is non-trivial.

\begin{figure}[htbp]
  \centering
  \includegraphics[width=\columnwidth]{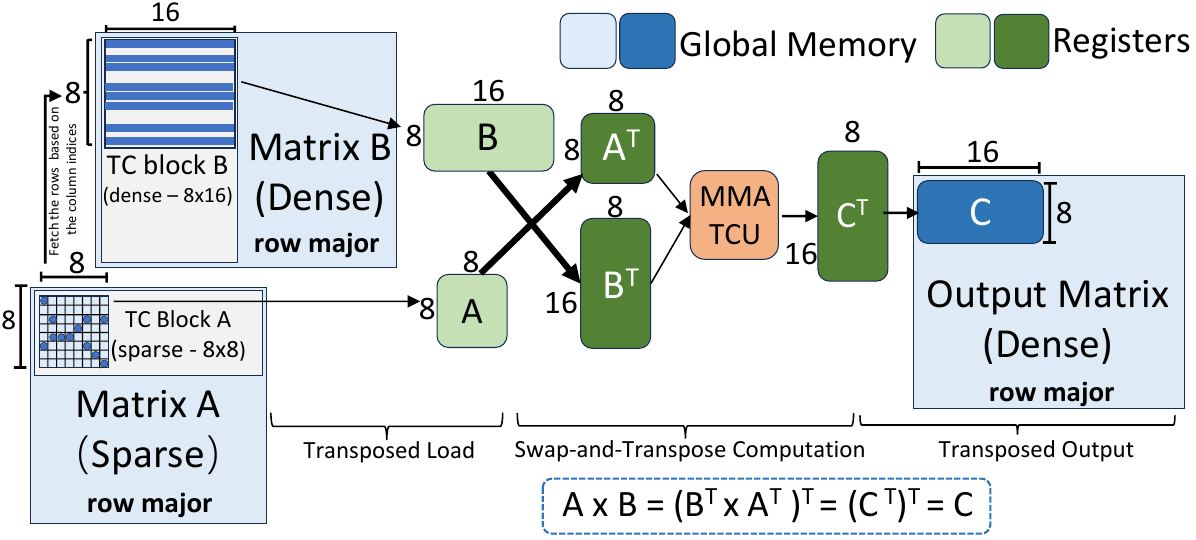}
  \caption{The implementation of SpMM with the swap-and-transpose MMA computation strategy.}
  \label{fig:spmm_trans}
\end{figure}

\subsection{The implementation of SpMM}

\begin{figure}[htbp]
  \centering
  \includegraphics[width=\columnwidth]{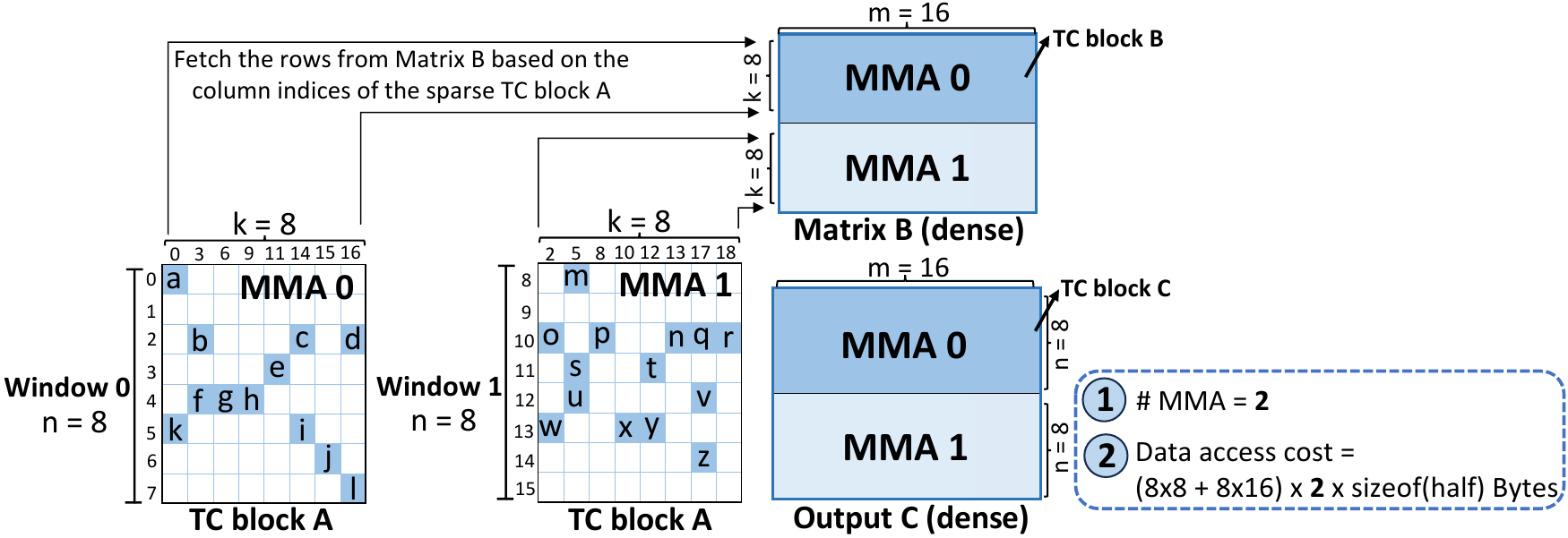}
    \caption{SpMM with 8$\times$1 vector size in FlashSparse. The original sparse matrix is shown in Figure~\ref{fig:16times1} (a).}
    \label{fig:moti2}
\end{figure}

In this section, we detail the design of our SpMM kernel with the swap-and-transpose MMA computation strategy.
As shown in Figure~\ref{fig:spmm_trans}, the threads within a warp first load the sparse TC block A from original matrix A and dense TC block B from original matrix B.
Benefiting from the swap-and-transpose MMA computation, the shape of the sparse TC block A in FlashSparse is 8 $\times$ 8 instead of 16 $\times$ 8 (current SOTA), while the shape of the dense TC block B is 8 $\times$ 16. To visually show the benefit of the swap-and-transpose MMA computation in FlashSparse, we use the same sparse matrix A in Figure~\ref{fig:16times1} (a) to compute SpMM as an example. Figure~\ref{fig:moti2} shows that using the 8$\times$1 vector size to partition the sparse matrix only requires 2 MMAs to finish the SpMM in FlashSparse. In contrast, the current SOTA works using the 16$\times$1 vector size requires 4 MMAs as shown in Figure~\ref{fig:16times1} (b). 
As visualized in Figure~\ref{fig:moti2}, the sparse TC block A with shape 8$\times$8 is much denser than the 16$\times$8 block in Figure~\ref{fig:16times1} (b), which demonstrates that the 8$\times$1 vector size helps to reduce the number of zeros in the sparse TC block and the computation redundancy. 
Additionally, the data access cost in Figure~\ref{fig:16times1} (b) is also proportionally reduced by 50\% in Figure~\ref{fig:moti2}.
Therefore, the swap-and-transpose strategy which enables a smaller vector size is superior for both computation and data access efficiency compared to SOTAs.

\begin{figure*}[htbp]
  \centering
  \includegraphics[width=18cm]{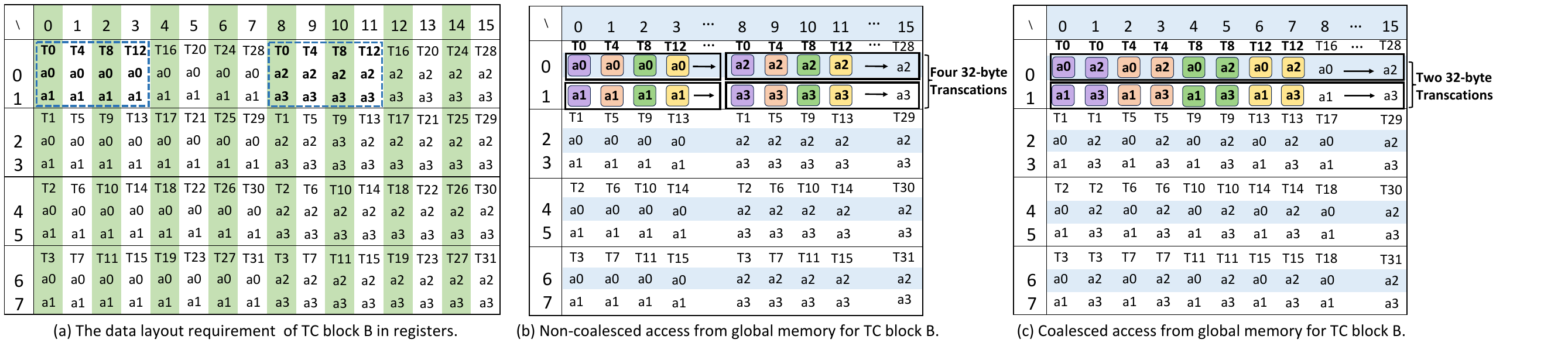}
  \caption{Memory-efficient thread mapping for coalesced data access. The precision of each element is FP16.}
  \label{fig:shuffle}
\end{figure*}

Furthermore, implementing the swap-and-transpose MMA strategy for SpMM requires addressing the issue of data access efficiency.
As shown in Figure~\ref{fig:transpose}, the data layout requirement of the sparse TC block A is row-major, whereas the dense TC block B is column-major.
The original sparse matrix A can be pre-stored in row-major format (as illustrated in Section 3.5).
However, the TC block B is formed by the rows of the dense matrix B indexed by the column indices of nonzero vectors in the TC block A, which are noncontinuous in memory address. To efficiently load the data of matrix B to registers is challenging.

As shown in Figure~\ref{fig:shuffle} (a), the data layout of TC block B is consistent with the transposed data layout of the left operand of MMA presented in the official technical documentation~\cite{m16n8k8}. 
Figure~\ref{fig:shuffle} (b) shows the direct mapping of threads to global memory according to the data layout shown in Figure~\ref{fig:shuffle} (a), and each thread needs to load four elements with precision FP16 (i.e. a0, a1, a2 and a3) from global memory (row-major) to registers.
However, these four elements are located in global memory with large strides, leading to low memory access efficiency. For example, for each element access, the data accessed by \{T0, T4,...,T28\} (a group with 8 threads) form a data block of 16 bytes, which is still less than the minimum memory transaction size (i.e., 32 bytes) supported on NVIDIA GPUs. Note that NVIDIA GPUs support three memory transaction sizes, including 32 bytes, 64 bytes, and 128 bytes. This means that although only 16 bytes data are accessed, a 32-byte memory transaction has to be transferred. As a result, the direct mapping strategy shown in Figure~\ref{fig:shuffle} (b) requires total 16 memory transactions to access the entire TC block B from global memory, leading to a waste of 50\% data movement.

Since the data layout requirements for TC block B and C in Figure~\ref{fig:transpose} are consistent, the data locations accessed by each thread for computation are aligned with the locations where the results are stored.
This alignment allows us to directly swap the columns of TC Block B in the registers.
Therefore, we propose a memory-efficient thread mapping strategy which enables coalesced data access, as shown in Figure~\ref{fig:shuffle} (c). 
The key idea is to shuffle the columns that threads need to access, the effectiveness of which is to let the four FP16 elements accessed by each thread form a 2$\times$2 block. For example, in Figure~\ref{fig:shuffle} (b), thread 0 (T0) is responsible for accessing the four elements from 0th and 8th columns of TC block B; in contrast, in Figure~\ref{fig:shuffle} (c), T0 accesses the four elements from the adjacent 0th and 1st columns. In the 2$\times$2 block, two FP16 elements in each row are access using a single element in FP32 data type.
Taking \{T0, T4,...,T28\} (a group with 8 threads) as an example, all the 16 FP16 elements in each row accessed by these 8 threads can be coalesced into a single 32-byte memory transaction, matching the minimum memory transaction granularity. The same applies to the other thread groups in a warp. Therefore, in Figure~\ref{fig:shuffle} (c), accessing all the elements of TC block B requires only 8 32-byte memory transactions (50\% reduction compared to the direct mapping in Figure~\ref{fig:shuffle} (b)), which is important for memory-bound sparse operators.
Additionally, the final output result C$^{T}$ still needs to be transposed and stored back to the global memory. 
Since the data layouts of B$^{T}$ and C$^{T}$ are identical on the registers of the threads in a warp, the final output results can be written back to global memory in a similar way to the data access of TC block B, enabling efficient coalesced data write back.

\subsection{The implementation of SDDMM}
\begin{figure}[htbp]
  \centering
  \includegraphics[width=\columnwidth]{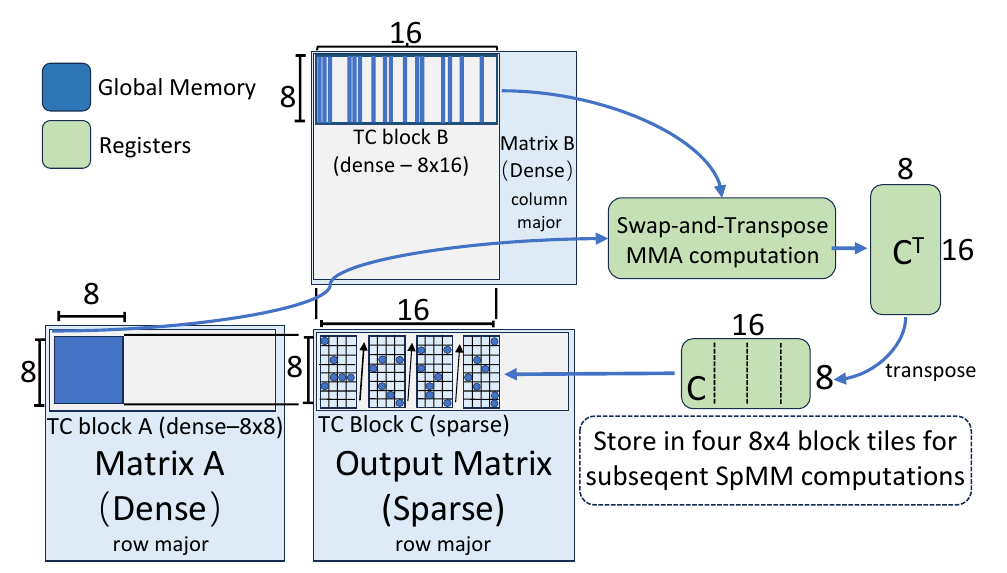}
  \caption{The implementation of SDDMM based on the swap-and-transpose MMA computation strategy.}
  \vspace{-1em}
  \label{fig:sddmm_trans}
\end{figure}

SDDMM is another major sparse operator in which the two input matrices are dense while the output matrix is sparse by sampling. 
In various fields, the sparse output matrix  from SDDMM serves as the input matrix for SpMM.
For example, in the attention-based GNNs~\cite{thekumparampil2018attention,niu2022tilespgemm,tylenda2009towards,kunegis2009learning}, a sparse attention matrix is usually computed first using SDDMM, and then aggregate with the feature matrix using SpMM. The sparse output matrix of SDDMM usually has high sparsity and irregularity. The swap-and-transpose strategy also works for SDDMM to significantly reduce the computation redundancy by enabling a smaller vector size for the sparse output matrix. As shown in the Figure~\ref{fig:sddmm_trans}, by benefiting from the swap-and-transpose MMA computation strategy, the sparse TC block C in our SDDMM kernel is 8$\times$16 (FlashSparse with TF32 and FP16) instead of 16$\times$8 (SOTA with TF32). By using a smaller vector size of 8$\times$1, the sparse TC block C is denser than that using a vector size of 16$\times$1. Besides, the matrix A is row-major while the matrix B is column-major, which aligns perfectly with the required data layout of the swap-and-transpose MMA computation.

\begin{figure}[htbp]
  \centering
  \includegraphics[width=\columnwidth]{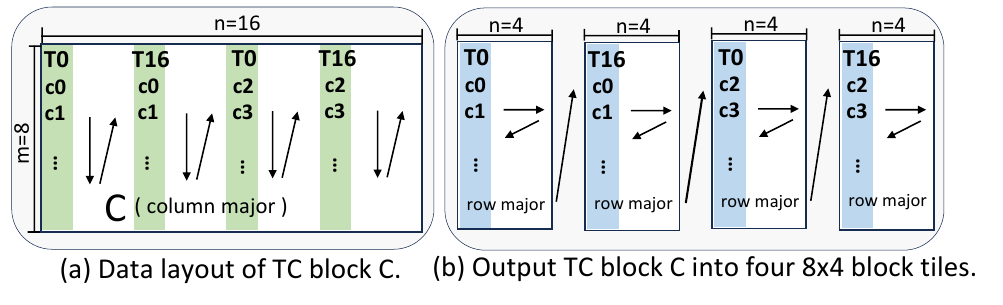}
  \caption{Output splitting in SDDMM for subsequent SpMM.}
  \label{fig:output}
\end{figure}

However, the shapes of the sparse TC block A in SpMM are 8$\times$8 and 8$\times$4 for FP16 and TF32, respectively.
Therefore, as illustrated in Figure~\ref{fig:output}, we split the sparse TC block C into 4 sub-blocks for storage to accommodate the format of the subsequent SpMM computations (TF32).
Besides, the data layout requirement of output C in the swap-and-transpose MMA computation is column-major, while the storage format of TC block A in SpMM is row-major (as shown in Figure~\ref{fig:transpose}).
Therefore, it is not easy to compute the target write positions for each thread.
Algorithm~\ref{algo:output} is responsible for calculating the positions of target elements in the output matrix C (global memory).
Specifically, 
we calculate the target position of c0 in the sparse matrix C based on $tid$ (lines 2-8). 
Finally, starting from the target position, iteratively write c0, c1, c2, and c4 into the sparse matrix C (lines 9-15).

\begin{algorithm}
  \caption{Output splitting in SDDMM}\label{algo:output}
  \small
  \begin{algorithmic}[1]
    \Statex \textbf{Input:} The c0,c1,c2,c3 of each thread in $C^{T}$
    \Statex \textbf{Output:} Sparse Matrix C
    \State $tid \gets threadId.x \% 32$
    \State \textcolor{blue}{// The target offset for 8$\times$8 sub-blocks (FP16) in the Matrix C}
    \State $offset\_8\_8 \gets (tid \% 4) * 2 * 8 + (tid/4)$
    \State \textcolor{blue}{// The target offset for 8$\times$4 sub-blocks (TF32) in the Matrix C}
    \If{$tid > 15$} {$k=1$} 
    \Else{ $k=0$ }
    \EndIf
    \State $offset\_8\_4 \gets (tid \% 4) * 2 * 4 + (tid/4) + (k*32) - (k*4) $
    \State \textcolor{blue}{// Write the c0,c1,c2,c3 from register to global memory}
    \State \textcolor{blue}{// Taking $offset\_8\_4$ as an example}
    \State $C \gets C + offset\_8\_4 $
    \For{$i \gets 0$ to $1$}  
            \State $C + i*64 \gets c\_(2*i)$ \textcolor{blue}{// If this position not zero}
            \State $C + i*64 + 4 \gets c\_(2*i+1)$ \textcolor{blue}{// If this position not zero}
    \EndFor
  \end{algorithmic}
\end{algorithm}

\subsection{Memory efficient storage format}
In collaborating with the swap-and-transpose MMA computation for sparse operators, an efficient sparse matrix storage format is indispensable.
Under the constraints of MMA's strict operand shape, the number of vectors in each window should be an integer multiple of $k$. Existing work~\cite{magicube} resolves this issue by padding zero vectors, but this results in high memory overhead especially for highly sparse matrices.
However, we observe that zero vector padding occurs only in the last TC block of each window. 
The number of padded zero vectors can be identified using a modulo operation on the kernel side.
Therefore, we propose a TCU-friendly and memory-efficient storage format, ME-BCRS, which stores only nonzero vectors without zero vector padding.

\begin{figure}[htbp]
  \centering
  \includegraphics[width=\columnwidth]{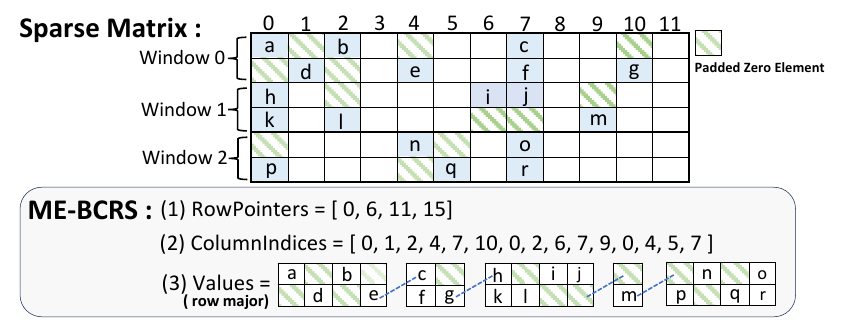}
  \caption{ME-BCRS storage format.}
  \label{fig:format}
\end{figure}

As shown in the Figure~\ref{fig:format}, ME-BCRS utilizes three arrays to represent the sparse TC blocks within each window.
For ease of illustration, we use a sparse TC block shape of 2$\times$4 as an example. 
\ding{172} $RowPointers$ indicates the starting index of each row window in the $ColumnIndices$.
\ding{173} $ColumnIndices$ holds the column indices of non-zero vectors in each sparse TC block.
\ding{174} $Values$ uses sparse TC blocks as strides, storing the elements of each sparse TC block in row-major to meet the data layout requirement of TC block A in swap-and-transpose MMA.
Since the ME-BCRS format does not store these padded zero vectors, the column dimension of TC blocks in ME-BCRS varies but does not exceed k, as shown in Figure~\ref{fig:format}.
Besides, taking the memory space for $RowPointers$ as an example, we only need to store M row pointers (M is the number of row windows of a sparse matrix) rather than 2M in the padding-based scheme~\cite{magicube}.
This is because we only record information for nonzero vectors. 
Thus, we need to implement specific SpMM and SDDMM algorithms on the kernel side to compute the last TC block A in each window.

Taking SpMM as an example, each thread in a warp first uses a modulo operation to compute the number of residue vectors as $residue$ in the last TC block A within each window.
Next, we calculate $column\_offset$ within the last TC block A that the current thread needs to access.
If $column\_offset$ is greater than the $residue$, it means that the vector indicated by $column\_offset$ belongs to the next window.
In this case, we set the register values provided the current thread for TC blocks to 0. 
Otherwise, these values need to be accessed from global memory using $column\_offset$. 
Overall, ME-BCRS effectively reduces memory footprint of the sparse matrix storage format through padding elimination and efficient kernel implementation.

\section{Evaluation}
\textbf{Benchmarks and Platforms: }We incorporate FlashSparse into the Pytorch framework for performance evaluation.
The swap-and-transpose is built into the sparse kernels (both SpMM and SDDMM).
Our experimental platform consists of a NVIDIA H100 GPU and a GeForce RTX4090 GPU. 
1) \textbf{NVIDIA H100 PCIe} has 456 Tensor Core unites and 14592 CUDA cores, 80 GB graphics memory.
2)\textbf{NVIDIA GeForce RTX4090} has 512 Tensor Core unites and 16384 CUDA cores, 24 GB graphics memory.

\textbf{Baselines: }We compare FlashSparse with state-of-the-art approaches for sparse operators on GPUs and end-to-end frameworks. 
First, for approaches on CUDA cores:
\ding{172} \textbf{RoDe}~\cite{RoDe}, a row decomposition based approach to optimize the SpMM and SDDMM kernels on GPUs.
\ding{173} \textbf{Sputnik}~\cite{sputnik}, a one-dimensional tiling and rotation technique to address the load imbalance issue for sparse kernels.
\ding{174} \textbf{GNNAdvisor}~\cite{wang2021gnnadvisor}, an efficient runtime system to accelerate various sparse workloads by introducing 2D workload management.
\ding{175} \textbf{GE-SpMM}~\cite{huang2020ge} introduces the Coalesced Row Caching (CRC) method for SpMM, which uses GPU shared memory to cache sparse matrix rows.
\ding{176} \textbf{cuSPARSE}~\cite{cuSPARSE} is a library developed by NVIDIA for performing efficient sparse matrix operations.
Additionally, for works on Tensor Cores:
\ding{177} \textbf{DTC-SpMM}~\cite{dtc} is a novel approach with systematic optimizations for accelerating general SpMM on TCUs.
\ding{178} \textbf{TC-GNN}~\cite{wang2023tc} accelerates GNN training with WMMA instructions on TCUs.
Moreover, for end-to-end GNNs frameworks:
\ding{179} Deep Graph Library (\textbf{DGL})~\cite{dgl} is a widely used and well-maintained GNNs framework with the support for high-performance sparse matrix computation.
\ding{180} PyTorch Geometric (\textbf{PyG})~\cite{fey2019fast} is another popular GNNs framework, which is based on edge-wise parallelization.

\newcommand{\cmark}{\color{blue}\ding{51}}
\newcommand{\xmark}{\color{red}\ding{55}}
\setlength{\tabcolsep}{2pt}
\begin{table}[!ht]
    \centering
    \caption{The precision supported in the Baselines and FlashSparse.}
    \begin{tabular}{ccccc}
    \toprule
     \multirow{2}[2]{*}{Baselines} & \multicolumn{3}{c}{Precision} & \multirow{2}[2]{*}{Granularity} \\
     \cmidrule(lr){2-4}
     & FP32 & TF32 & FP16 & \\
    \midrule
    RoDe  & \cmark & \xmark &  \xmark & CUDA cores \\
    Sputnik  & \cmark & \xmark & \xmark  & CUDA cores  \\
    GNNAdvisor & \cmark & \xmark &  \xmark & CUDA cores \\
    GE-SpMM & \cmark & \xmark &  \xmark & CUDA cores \\
    cuSPARSE  & \cmark & \xmark & \xmark & CUDA cores  \\
    DGL & \cmark & \xmark &  \xmark & CUDA cores  \\
    PyG  & \cmark & \xmark &  \xmark & CUDA cores \\
    DTC-SpMM & \xmark & \cmark &  \xmark & 16$\times$1 on TCU \\
    TC-GNN &\xmark & \cmark & \xmark & 16$\times$1 on TCU\\
    {\bf FlashSparse} & \xmark & \cmark & \cmark  & {\bf8$\times$1 on TCU} \\
    \bottomrule
    \end{tabular}
    \label{tab:percision}
\end{table}

\begin{table}[!ht]
    \centering
    \caption{Graph datasets for evaluation.}
    \begin{tabular}{cccc}
    \toprule
    Dataset & \#Vertex &  \#Edge &  AvgRowLength\\
    \midrule
    GitHub & 37,700 & 615,706 & 16.33\\ 
    Artist & 50,515 & 1,638,396  & 32.4\\ 
    Blog & 88,784 & 4,186,390 & 47.2 \\ 
     Ell & 203,769 & 672,479 & 3.3\\ 
     Yelp & 716,847 & 13,954,819 & 19.46\\ 
     DD & 334,925 & 1,686,092 & 5.03\\ 
    Reddit & 232,965 & 114,848,857 & 492.98\\ 
     Amazon & 403,394 & 9,068,096 & 22.48\\ 
     Amazon0505 & 410,236 & 4,878,874 & 11.89\\ 
     Comamazon & 334,863 & 1,851,744 & 5.5\\ 
     Yeast & 1710902 & 5347448 & 3.1 \\
    OGBProducts & 2,449,029 & 126,167,053  & 51.52\\ 
    AmazonProducts & 1,569,960 & 264,339,468 & 128.37\\ 
     IGB-small & 1,000,000 & 13,068,130 & 13.06\\
    IGB-medium & 10,000,000 & 129,994,908 & 12.99\\ 

    \bottomrule
    \end{tabular}
        \label{tab:dataset}
\end{table}

We summarize the precision types supported by all the baselines in Table~\ref{tab:percision}.
The baselines on CUDA cores utilize FP32, while those on TCUs employ TF32. 
FlashSparse supports both TF32 and FP16 on TCUs.
Additionally, for the tunable baseline, its optimal version is used in evaluation. 
For DGL and PyG, we use their latest open-source versions as strong baselines.

\textbf{Datasets: } 
We select sparse matrices from the SuiteSparse collection~\cite{SuiteSparse} that have more than 10k rows, 10k columns, and 100K nonzeros, which is consistent with the criteria used in RoDe.
Since TC-GNN can only handle square matrices~\cite{dtc}, we finally choose a representative subset of 500 matrices.
In addition to the SuiteSparse matrices, we extend our evaluation to include matrices from GNNs.
We choose 15 classical graph datasets from real-world applications, such as IGB~\cite{khatua2023igb}, AmazonProducts~\cite{zeng2019graphsaint} (as shown in Table~\ref{tab:dataset}). Overall, total 515 different sparse matrices are used for evaluation.


\begin{figure*}[htbp]
  \centering
  \includegraphics[width=0.95\linewidth]{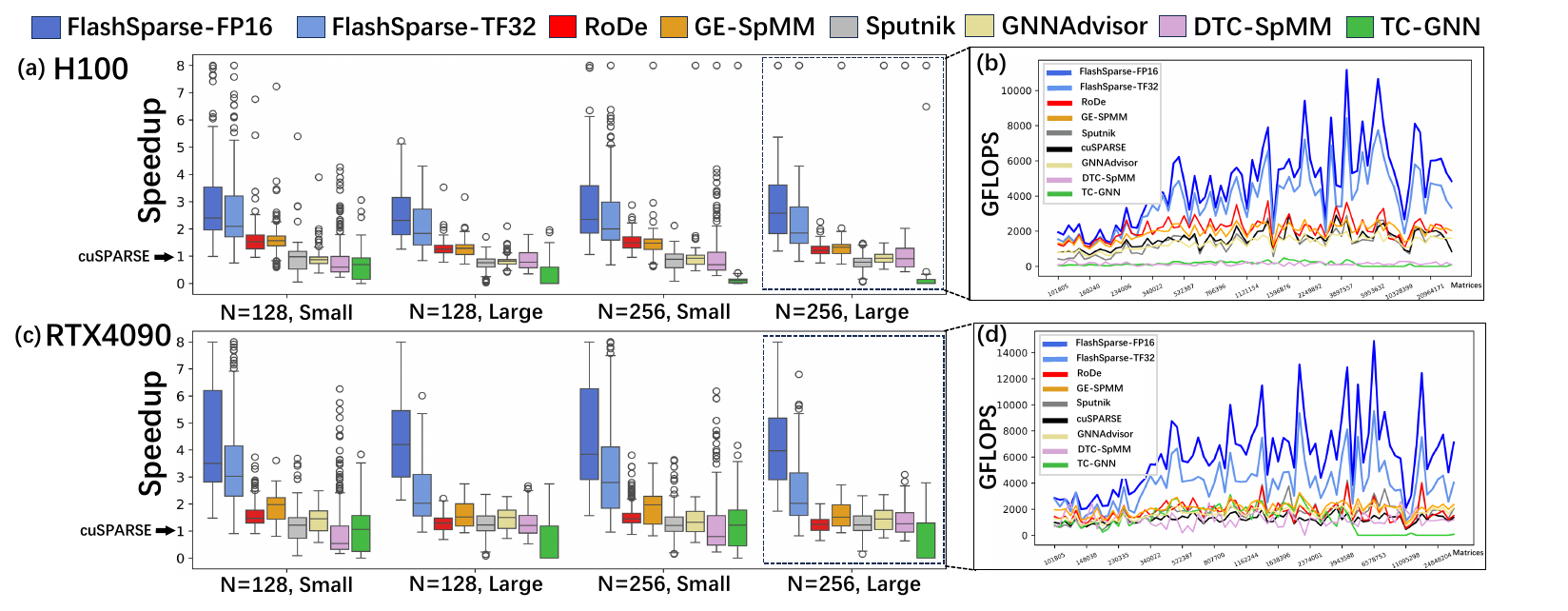}
    \caption{SpMM performance results on H100 and RTX4090 GPU. (a)(c) The speedup distribution of FlashSparse and baselines over cuSPARSE. (b)(d) The measured throughput of the 515 matrices with N=256. The matrices are sorted in ascending order according to the number of nonzero elements, and each point represents the average GFLOPS of six consecutive matrices.}
    \label{fig:spmm}
\end{figure*}

\begin{table*}[]
    \centering
    \caption{Detailed Speedup distribution of SpMM for FlashSparse over baselines on H100 and RTX4090 GPUs (N=128).}
    \begin{tabular}{ccccccccccccc}
    \toprule
& \multicolumn{5}{c}{H100} &&& \multicolumn{5}{c}{RTX4090} \\
\cmidrule(lr){2-6} \cmidrule(lr){9-13} 
{\bf Speedup} & TC-GNN & DTC-SpMM & RoDe &  Sputnik & GE-SpMM    &&& TC-GNN & DTC-SpMM & RoDe & Sputnik  & GE-SpMM\\
\midrule


<1 & 0.0\% & 0.0\% & 2.38\% & 0.2\% & 1.19\% &&& 0.0\% & 0.0\% & 0.2\% & 0.0\% & 0.2\%   \\ 
1-1.5 & 0.0\% & 0.0\% & 23.21\% & 1.39\% & 18.25\% &&& 1.6\% & 0.2\% & 5.81\% & 0.4\% & 9.82\%   \\ 
1.5-2 & 0.99\% & 0.99\% & 24.21\% & 2.98\% & 21.63\% &&& 2.61\% & 1.0\% & 13.83\% & 5.01\% & 19.04\%   \\ 
$\geq2$ & 99.01\% & 99.01\% & 50.2\% & 95.43\% & 58.93\% &&& 95.79\% & 98.8\% & 80.16\% & 94.59\% & 70.94\%   \\ 
{\bf Mean} & $\geq50$ &4.41x &2.23x &4.42x &2.25x &&& $\geq50$ &5.5x &3.22x &4.28x &2.67x  \\ 
{\bf Max} & $\geq50$ &16.03x &6.9x &52.95x &7.34x &&& $\geq50$ &25.26x &14.2x &43.53x &12.19x  \\ 
\bottomrule
    \end{tabular}
    \label{tab:spmm}
\end{table*}

\subsection{SpMM Evaluation}
We measure the SpMM performance with different settings for N (the number of columns in dense matrix B), including 128 and 256 (consistent with DTC-SpMM). 
Figure~\ref{fig:spmm} illustrates the speedup distribution and throughput (GFLOPS) of FlashSparse and baselines.
We category the sparse matrices into two groups based on the number of rows (one hundred thousand): small and large.
As depicted in the Figure~\ref{fig:spmm}(a)(c), the median speedup (normalized to cuSPARSE) of FlashSparse outperforms all the baselines in both FP16 and TF32 precisions across all settings. 
Note that, for clarity in displaying the distribution of baseline speedups, we cap the speedup of FlashSparse at 8x, meaning the actual median speedup of FlashSparse could be higher.
Meanwhile, Figures~\ref{fig:spmm}(b) and (d) shows that FlashSparse also achieves the highest computational throughput.
TC-GNN performs extremely poorly on matrices with more than 5 million nonzero elements, so we label its GFLOPS as 0.
This is because TC-GNN uses a 16$\times$1 vector granularity and its algorithm design requires extensive position checks for sparse elements within the kernel.
This overhead becomes more pronounced with larger matrices. 
In addition, the performance gap between TCUs and CUDA cores is larger on the RTX4090 than on the H100.
Thus, the throughput gap between FlashSparse and RoDe on RTX4090 is more pronounced. 
Overall, the geometric mean throughput of FlashSparse is 4888 GFLOPS (up to 26 TFLOPS) for FP16 precision and 2697 GFLOPS (up to 16 TFLOPS) for TF32 precision on RTX4090 GPU.

\begin{figure}[htbp]
  \centering
  \includegraphics[width=\columnwidth]{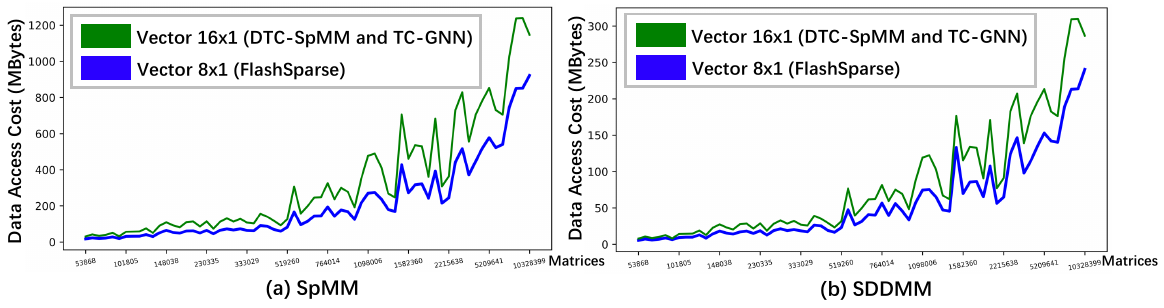}
  \caption{The data access cost of 16$\times$1 and 8$\times$1 vectors in SpMM (N=128) and SDDMM (N=32). The precision is FP16.}
  \vspace{-1em}
  \label{fig:mem_all}
\end{figure}

\begin{figure*}[htbp]
  \centering
  \includegraphics[width=18cm]{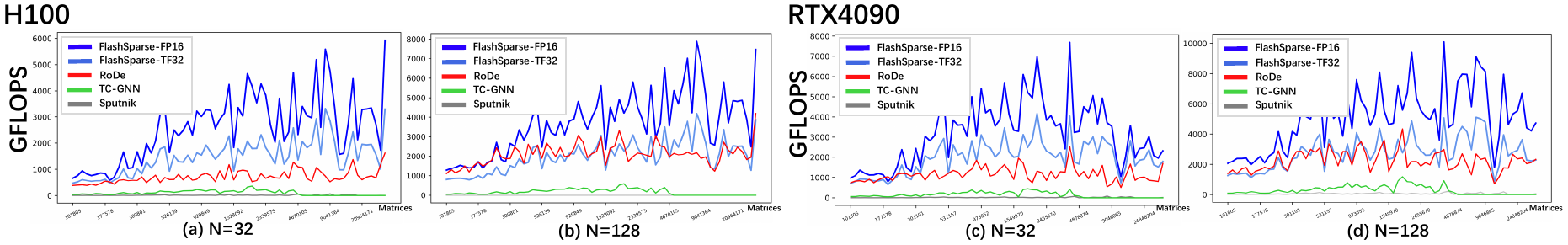}
    \caption{SDDMM performance on H100 and RTX4090 GPUs. The matrices are sorted in a manner consistent with Figure~\ref{fig:spmm}.}
    \label{fig:sddmm}
\end{figure*}

Table~\ref{tab:spmm} presents the speedup distribution of Figure~\ref{fig:spmm}. 
The experimental results show that FlashSparse achieves geometric mean speedups of 5.5x (up to 25.26x) and 3.22x (up to 14.2x) over DTC-SpMM (SOTA on TCUs) and RoDe (SOTA on CUDA cores) respectively on RTX4090 GPU.
There are several reasons why FlashSparse maintains the highest speedup.
First, TCUs provide significantly higher peak computational performance compared to CUDA cores. 
For example, the peak performance of TCUs with FP16 on the H100 GPU is 30 times higher than that of CUDA cores with FP32. 
This superior matrix arithmetic capability highlights the potential of TCUs for accelerating sparse operators. 
However, directly applying the dense computation units to sparse arithmetic can lead to severe computation and data access redundancy.
The key to addressing this issue lies in finer vector granularity.
DTC-SpMM and TC-GNN are both limited to a 16$\times$1 vector granularity, whereas our swap-and-transpose MMA strategy enables FlashSparse with a finer 8$\times$1 vector granularity.
Figure~\ref{fig:mem_all} (a) presents the data access cost of 515 sparse matrices to complete one SpMM computation in FlashSparse (N=128).
The data access cost refers to the cost of loading data from the memory hierarchy, without distinguishing the data sources (from global memory or caches).
As shown in Figure~\ref{fig:mem_all} (a), 8$\times$1 vector size can reduce the data access cost by up to 49\% (averaged 35\%) compared to 16$\times$1 vector size. In addition, the computation cost is also reduced proportionally.




\subsection{SDDMM Evaluation}
Figure~\ref{fig:sddmm} and Table~\ref{tab:sddmm} illustrate the SDDMM performance comparison and speedup distribution respectively. 
The settings for N are 32 and 128, consistent with RoDe.
The experimental results demonstrate that both the throughput and speedup of FlashSparse are notable.
Specifically, FlashSparse achieves geometric mean SDDMM speedups of 2.92x (up to 18.59x) and 2.18x (up to 14.93x) over the state-of-the-art work RoDe on H100 and RTX4090 GPU, respectively.
Moreover, the main reason for the poor performance of TC-GNN is also the usage of 16$\times$1 vector granularity and position checks we mentioned above. 
We also calculate the data access cost for completing one SDDMM computation in FlashSparse (N = 32).
As show in Figure~\ref{fig:mem_all} (b), the 8$\times$1 vector granularity can reduce the data access cost by up to 49\% (with an average reduction of 28\%) compared to the 16$\times$1 vector granularity.

\begin{table}[]
    \centering
    \caption{Detailed speedup distribution of SDDMM on H100 and RTX4090 GPUs (N=32).}
    \begin{tabular}{ccccc}
    \toprule
& \multicolumn{2}{c}{H100} & \multicolumn{2}{c}{RTX4090} \\
\cmidrule(lr){2-3} \cmidrule(lr){4-5} 
{\bf Speedup} & TC-GNN  & RoDe & TC-GNN  & RoDe\\
\midrule
<1 & 0.0\% & 13.63\% & 0.0\% & 21.4\%   \\ 
1-1.5 & 0.0\% & 10.06\% & 0.0\% & 13.56\%   \\ 
1.5-2 & 0.0\% & 7.97\% & 0.0\% & 15.25\%   \\ 
$\geq2$ & 100.0\% & 68.34\% & 100.0\% & 49.79\%   \\ 
{\bf Geo Mean} & 100.0x &2.92x &100.0x &2.18x  \\ 
{\bf Max} & 100.0x &18.59x &100.0x &14.93x  \\ 
\bottomrule
    \end{tabular}
    \label{tab:sddmm}
    \vspace{-1em}
\end{table}



\subsection{Ablation Study}
We conduct ablation study on H100 and RTX4090 GPUs to validate the effectiveness of our optimization strategies.

\begin{figure}[htbp]
  \centering
  \includegraphics[width=\columnwidth]{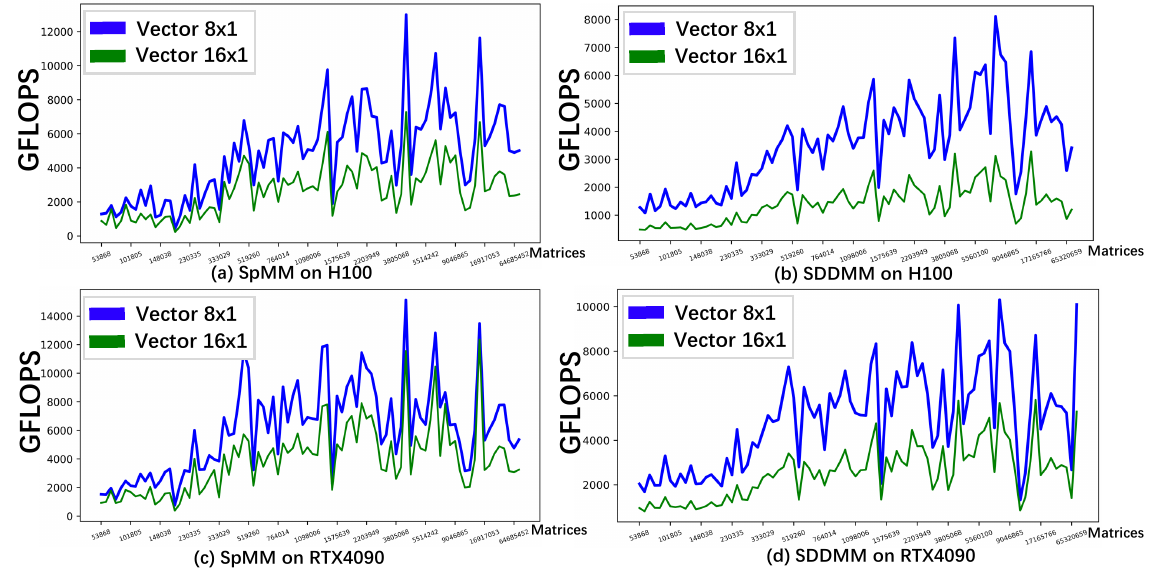}
  \caption{The throughput of FlashSparse under 16$\times$1 and 8$\times$1 vector granularity in SpMM and SDDMM.}
  \label{fig:ab1}
\end{figure}

\textbf{The swap-and-transpose MMA computation strategy.}
To verify the effectiveness of the 8$\times$1 vector size achieved by the swap-and-transpose strategy, we implement a baseline version with 16$\times$1 vector size (the other is the same as FlashSparse) for performance comparison.
We measure the throughput (GFLOPS) of FlashSparse (8$\times$1 vector size) and the baseline (16$\times$1 vector size) for SpMM and SDDMM.
As illustrated in Figure~\ref{fig:ab1}, FlashSparse with 8$\times$1 vector size outperforms 16$\times$1 in both SpMM and SDDMM across all 515 matrices. FlashSparse achieves the geometric mean speedup of 1.89x (up to 3.44x) for SpMM and 2.61x (up to 3.85x) for SDDMM on H100 compared to the 16$\times$1 version. 
The results demonstrate that the reduction of computation and data access redundancy achieved by our swap-and-transpose strategy brings practical performance improvement. This further confirms the importance of finer vector sizes in enhancing the performance of sparse operators on TCUs.

\begin{figure}[htbp]
  \centering
  \includegraphics[width=\columnwidth]{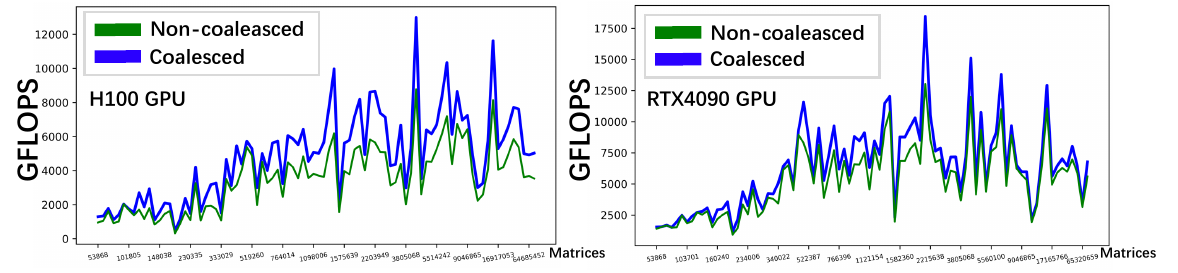}
  \caption{The throughput of FlashSparse under non-coalesced and coalesced data access modes in SpMM.}
  \label{fig:ab_access}
\end{figure}

\textbf{The memory-efficient thread mapping for coalesced data access.} 
We compare the performance of FlashSparse under non-coalesced (direct thread mapping) and coalesced (memory-efficient thread mapping) data access modes across 515 sparse matrices. As illustrated in Figure~\ref{fig:ab_access}, compared to the non-coalesced mode, the coalesced data access enabled by the memory-efficient thread mapping strategy achieves on average 1.34x (up to 2.0x) speedup on H100 and on average 1.18x (up to 2.0x) speedup on RTX4090.

\textbf{The effectiveness of ME-BCRS storage format.} We test 515 matrices to compare the memory footprint of ME-BCRS with SR-BCRS~\cite{magicube} (the zero vector padding-based method). 
As shown in the Table~\ref{tab:format}, ME-BCRS reduced the memory footprint occupied by the storage format by an average of 11.72\% (max 50.0\%) across 515 matrices, with 336 of them achieving a reduction of more than 10\%.

\begin{table}[h!]
    \centering
    \caption{The percentage of memory footprint reduced by ME-BCRS compared to SR-BCRS.}
    \begin{tabular}{cccccc}
    \toprule
    Percentage & 1\%-10\% & 11\%-20\% & 21\%-30\% & 31\%-40\% & $\geq41\%$  \\
    \midrule
    \#Matrices & 179 & 172 & 110 & 38 & 16 \\
    \bottomrule
    \end{tabular}
    \vspace{-1em}
    \label{tab:format}
\end{table}

\begin{figure*}[htbp]
  \centering
  \includegraphics[width=0.95\linewidth]{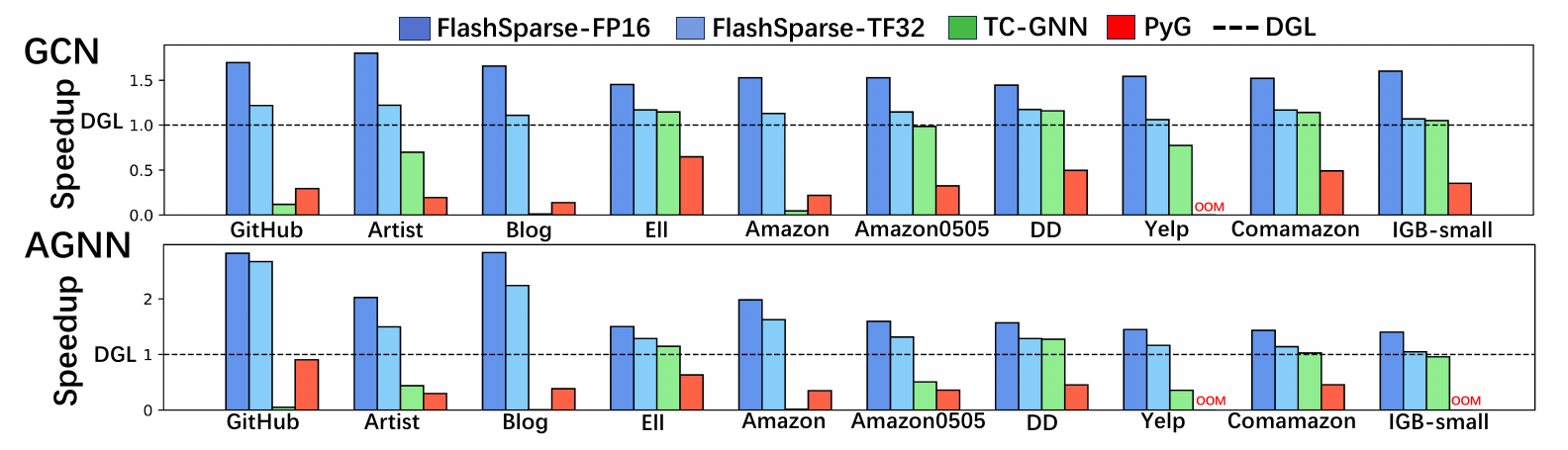}
    \caption{The end-to-end performance of FlashSparse on H100 and RTX4090 over GCN and AGNN models.}
    \label{fig:gnn}
\end{figure*}

\subsection{Case study: End-to-end GNNs performance}
We integrate FlashSparse into Pytorch framework and choose two popular GNN models, i.e. GCN~\cite{kipf2016semi,gibert2012graph,peng2024maxk} and AGNN~\cite{thekumparampil2018attention,niu2022tilespgemm,tylenda2009towards,kunegis2009learning} for end-to-end evaluation. 
GNN models consist primarily of feature aggregation (sparse operation) and feature update processes (dense operation):
\begin{align}
a_{v}^{(k+1)} &= \mathbf{Aggregate}((h_{u}^{(k)}, e_{u}^{(k)} | u \in \mathbf{N}(v)) \cup h_v^{(k)}) \\
h_{v}^{(k+1)} &= \mathbf{Update}(a_{v}^{(k+1)}, W) \label{equ:gnns}
\end{align}
where $u$ is the neighbor node of $v$ from $N(v)$; $h_{u}^{(k)}$ and $e_{u}^{(k)}$ are the feature vector and edge information for node $v$ at layer $k$; $a_{v}^{(k+1)}$ is the aggregated information of node $v$ in layer $k+1$; $W$ is the trainable weight matrix~\cite{wang2021gnnadvisor}.
GCN aggregates features of neighbor nodes by the SpMM operator. 
On the other hand, AGNN first computes attention for each edge using the SDDMM operator and subsequently aggregates node features using the SpMM operator.

The graph datasets are from various domains, as shown in Table~\ref{tab:dataset}.
We set the GNNs hidden layer dimensions to 128 for GCN and 32 for AGNN, respectively. 
The end-to-end time includes the times of format translation, forward and backward propagation through the models, and model update using gradients.
We select the latest high-performance versions of the DGL and PyG and the state-of-the-art GNNs frameworks using TCUs, TC-GNN~\cite{wang2023tc} for performance comparison.
Note that, TC-GNN does not set the softmax layer in AGNN model. 
But softmax layer still occupies a certain proportion of end-to-end time.
As shown in Figure~\ref{fig:gnn}, FlashSparse outperforms all baselines in both the GCN and AGNN models.
To summarize from Figure~\ref{fig:gnn}, compared with the latest version DGL, FlashSparse achieves the geometric mean speedup of 1.57x (up to 1.8x) for GCN and 1.79x (up to 2.83x) for AGNN on RTX4090 GPU.

Furthermore, in order to verify that FP16 and TF32 ensure comparable accuracy compared to FP32, we select several datasets collected by DGL to evaluate the end-to-end accuracy for GCN.
Accuracy refers to the Top-1 accuracy for node classification, the same as that in GCN~\cite{kipf2016semi} and AGNN~\cite{thekumparampil2018attention}.
We train 300 epochs of the GCN model with 5 layers.
As shown in Table~\ref{tab:acc}, the accuracy of GCN trained with TF32 and FP16 is comparable to that of GCN trained with FP32 using DGL and PyG (no accuracy loss).
In addition, we design GPU kernels to accelerate the conversion of the format from CSR to ME-BCRS. 
In static sparse scenarios, preprocessing only needs to be performed once. 
The preprocessing overhead accounts for only a tiny fraction of the end-to-end GNN runtime, approximately less than 1\%.

\section{Related Work}
The optimization and acceleration of SpMM and SDDMM have been the subject of extensive research~\cite{jiang2020novel,hong2019adaptive,magicube,chen2021efficient,wu2023turbomgnn,dakkak2019accelerating,feng2021apnn}.
The one-dimensional tiling scheme is first proposed by Sputnik~\cite{sputnik} for decomposing the sparse computation across processing elements.
In this scheme, each thread block computes a one-dimensional tiling of the output matrix. 
However, the distribution of nonzero elements in sparse matrices is irregular, leading to severe load imbalance. 
To solve this problem, RoDe~\cite{RoDe} proposed a 2D strategy to achieve workload balancing by dividing the rows of sparse matrices into regular and residue parts.
Besides, RoDe introduces new load balance and fine-grained pipeline technologies for further optimization.
Moreover, different from the programming flexibility of CUDA cores, sparse acceleration on TCUs is constrained by the strict data layout requirements of MMA~\cite{dasp,feldmann2021parallel,zachariadis2020accelerating}.
TC-GNN~\cite{wang2023tc} proposed the SGT technique, which divides the sparse matrix into nonzero vectors for computation on the TCU.
However, the current SOTA works on TCUs, TC-GNN and DTC-SpMM~\cite{dtc}, use the 16$\times$1 vector granularity by a simple adaptation to MMA instructions. This algorithm design philosophy imposes high redundancy for computation and data access, leading to low utilization of the TCUs’ computing power. Through sophisticated hardware-software co-design, Flash-Sparse achieves to use a more efficient nonzero vector size of 8$\times$1.
Compared to the SOTAs, our approach significantly reduces both computation and data access redundancy, and therefore brings unprecedented performance for sparse operators on TCUs.

\begin{table}[]
    \centering
    \caption{Accuracy of GCN with varying data precisions.}
    \begin{tabular}{cccccccc}
    \toprule
    \multirow{2}[2]{*}{Dataset} && PyG &&& DGL & FlashSparse & FlashSparse \\
    && FP32 &&& FP32 & FP16 & TF32 \\
    \midrule
    Cora && 71.2\% &&& 75.7\% & 75.1\% & 78.4\%  \\
    ELL && 95.12\% &&& 94.69\% & 95.18\% & 95.09\%  \\
    Pubmed && 75.6\% &&& 76.3\% & 77.3\% & 76.3\%  \\
    Questions && 97.03\% &&& 96.33\% & 97.01\% & 97.02\%  \\
    Minesweeper && 80.0\% &&& 79.96\% & 80.0\% & 80.0\%  \\
    \bottomrule
    \end{tabular}
    \label{tab:acc}
    \vspace{-1em}
\end{table}

\section{Conclusion}
In this paper, we propose FlashSparse, a novel approach that achieves the minimal 8$\times$1 vector granularity to accelerate SpMM and SDDMM. 
Its key innovation is to achieve the minimal vector granularity by swap-and-transpose MMA computation to reduce computation and data access redundancy of sparse operators on TCU.
In addition, FlashSparse is equipped with our proposed memory-efficient thread mapping strategy and a memory-efficient data format.
Our extensive experiments on H100 and RTX4090 GPUs show that FlashSparse sets new performance records for SpMM and SDDMM as well as end-to-end GNN performance. Our approach is also applicable to other TCU architectures with unbalanced matrix dimensions.

\section{Acknowledgement}
This project was supported by the National Natural Science Foundation of China under Grant No. 62372055, the National Science and Technology Major Project (2023ZD0120502), and the Fundamental Research Funds for the Central Universities.

\bibliographystyle{ACM-Reference-Format}
\bibliography{reference}


\begin{thebibliography}{48}


\ifx \showCODEN    \undefined \def \showCODEN     #1{\unskip}     \fi
\ifx \showDOI      \undefined \def \showDOI       #1{#1}\fi
\ifx \showISBNx    \undefined \def \showISBNx     #1{\unskip}     \fi
\ifx \showISBNxiii \undefined \def \showISBNxiii  #1{\unskip}     \fi
\ifx \showISSN     \undefined \def \showISSN      #1{\unskip}     \fi
\ifx \showLCCN     \undefined \def \showLCCN      #1{\unskip}     \fi
\ifx \shownote     \undefined \def \shownote      #1{#1}          \fi
\ifx \showarticletitle \undefined \def \showarticletitle #1{#1}   \fi
\ifx \showURL      \undefined \def \showURL       {\relax}        \fi
\providecommand\bibfield[2]{#2}
\providecommand\bibinfo[2]{#2}
\providecommand\natexlab[1]{#1}
\providecommand\showeprint[2][]{arXiv:#2}

\bibitem[ten(2023)]%
        {tensor}
 \bibinfo{year}{2023}\natexlab{}.
\newblock \bibinfo{title}{{Tensor Core. }}.
\newblock
\newblock
\newblock
\shownote{https://www.nvidia.cn/data-center/tensor-cores/}.


\bibitem[Almasan et~al\mbox{.}(2022)]%
        {almasan2022deep}
\bibfield{author}{\bibinfo{person}{Paul Almasan}, \bibinfo{person}{Jos{\'e} Su{\'a}rez-Varela}, \bibinfo{person}{Krzysztof Rusek}, \bibinfo{person}{Pere Barlet-Ros}, {and} \bibinfo{person}{Albert Cabellos-Aparicio}.} \bibinfo{year}{2022}\natexlab{}.
\newblock \showarticletitle{Deep reinforcement learning meets graph neural networks: Exploring a routing optimization use case}.
\newblock \bibinfo{journal}{\emph{Computer Communications}}  \bibinfo{volume}{196} (\bibinfo{year}{2022}), \bibinfo{pages}{184--194}.
\newblock


\bibitem[Anzt et~al\mbox{.}(2015)]%
        {anzt2015accelerating}
\bibfield{author}{\bibinfo{person}{Hartwig Anzt}, \bibinfo{person}{Stanimire Tomov}, {and} \bibinfo{person}{Jack~J Dongarra}.} \bibinfo{year}{2015}\natexlab{}.
\newblock \showarticletitle{Accelerating the LOBPCG method on GPUs using a blocked sparse matrix vector product.}. In \bibinfo{booktitle}{\emph{SpringSim (HPS)}}. \bibinfo{pages}{75--82}.
\newblock


\bibitem[Blei et~al\mbox{.}(2003)]%
        {blei2003latent}
\bibfield{author}{\bibinfo{person}{David~M Blei}, \bibinfo{person}{Andrew~Y Ng}, {and} \bibinfo{person}{Michael~I Jordan}.} \bibinfo{year}{2003}\natexlab{}.
\newblock \showarticletitle{Latent dirichlet allocation}.
\newblock \bibinfo{journal}{\emph{Journal of machine Learning research}} \bibinfo{volume}{3}, \bibinfo{number}{Jan} (\bibinfo{year}{2003}), \bibinfo{pages}{993--1022}.
\newblock


\bibitem[Bui et~al\mbox{.}(2022)]%
        {bui2022spatial}
\bibfield{author}{\bibinfo{person}{Khac-Hoai~Nam Bui}, \bibinfo{person}{Jiho Cho}, {and} \bibinfo{person}{Hongsuk Yi}.} \bibinfo{year}{2022}\natexlab{}.
\newblock \showarticletitle{Spatial-temporal graph neural network for traffic forecasting: An overview and open research issues}.
\newblock \bibinfo{journal}{\emph{Applied Intelligence}} \bibinfo{volume}{52}, \bibinfo{number}{3} (\bibinfo{year}{2022}), \bibinfo{pages}{2763--2774}.
\newblock


\bibitem[Chen et~al\mbox{.}(2021)]%
        {chen2021efficient}
\bibfield{author}{\bibinfo{person}{Zhaodong Chen}, \bibinfo{person}{Zheng Qu}, \bibinfo{person}{Liu Liu}, \bibinfo{person}{Yufei Ding}, {and} \bibinfo{person}{Yuan Xie}.} \bibinfo{year}{2021}\natexlab{}.
\newblock \showarticletitle{Efficient tensor core-based gpu kernels for structured sparsity under reduced precision}. In \bibinfo{booktitle}{\emph{Proceedings of the International Conference for High Performance Computing, Networking, Storage and Analysis(SC)}}. \bibinfo{pages}{1--14}.
\newblock


\bibitem[Dakkak et~al\mbox{.}(2019)]%
        {dakkak2019accelerating}
\bibfield{author}{\bibinfo{person}{Abdul Dakkak}, \bibinfo{person}{Cheng Li}, \bibinfo{person}{Jinjun Xiong}, \bibinfo{person}{Isaac Gelado}, {and} \bibinfo{person}{Wen-mei Hwu}.} \bibinfo{year}{2019}\natexlab{}.
\newblock \showarticletitle{Accelerating reduction and scan using tensor core units}. In \bibinfo{booktitle}{\emph{Proceedings of the ACM International Conference on Supercomputing}}. \bibinfo{pages}{46--57}.
\newblock


\bibitem[Davis and Hu(2011)]%
        {SuiteSparse}
\bibfield{author}{\bibinfo{person}{Timothy~A Davis} {and} \bibinfo{person}{Yifan Hu}.} \bibinfo{year}{2011}\natexlab{}.
\newblock \showarticletitle{The University of Florida sparse matrix collection}.
\newblock \bibinfo{journal}{\emph{ACM Transactions on Mathematical Software (TOMS)}} \bibinfo{volume}{38}, \bibinfo{number}{1} (\bibinfo{year}{2011}), \bibinfo{pages}{1--25}.
\newblock


\bibitem[{dgl}(2018)]%
        {dgl}
\bibfield{author}{\bibinfo{person}{{dgl}}.} \bibinfo{year}{2018}\natexlab{}.
\newblock \bibinfo{title}{{DGL}}.
\newblock
\newblock
\newblock
\shownote{https://docs.dgl.ai}.


\bibitem[Fan et~al\mbox{.}(2024)]%
        {dtc}
\bibfield{author}{\bibinfo{person}{Ruibo Fan}, \bibinfo{person}{Wei Wang}, {and} \bibinfo{person}{Xiaowen Chu}.} \bibinfo{year}{2024}\natexlab{}.
\newblock \showarticletitle{DTC-SpMM: Bridging the Gap in Accelerating General Sparse Matrix Multiplication with Tensor Cores}. In \bibinfo{booktitle}{\emph{ASPLOS24: Proceedings of the 29th ACM International Conference on Architectural Support for Programming Languages and Operating Systems, Volume 3}}. \bibinfo{pages}{253--267}.
\newblock


\bibitem[Feldmann et~al\mbox{.}(2021)]%
        {feldmann2021parallel}
\bibfield{author}{\bibinfo{person}{Johannes Feldmann}, \bibinfo{person}{Nathan Youngblood}, \bibinfo{person}{Maxim Karpov}, \bibinfo{person}{Helge Gehring}, \bibinfo{person}{Xuan Li}, \bibinfo{person}{Maik Stappers}, \bibinfo{person}{Manuel Le~Gallo}, \bibinfo{person}{Xin Fu}, \bibinfo{person}{Anton Lukashchuk}, \bibinfo{person}{Arslan~Sajid Raja}, {et~al\mbox{.}}} \bibinfo{year}{2021}\natexlab{}.
\newblock \showarticletitle{Parallel convolutional processing using an integrated photonic tensor core}.
\newblock \bibinfo{journal}{\emph{Nature}} \bibinfo{volume}{589}, \bibinfo{number}{7840} (\bibinfo{year}{2021}), \bibinfo{pages}{52--58}.
\newblock


\bibitem[Feng et~al\mbox{.}(2021)]%
        {feng2021apnn}
\bibfield{author}{\bibinfo{person}{Boyuan Feng}, \bibinfo{person}{Yuke Wang}, \bibinfo{person}{Tong Geng}, \bibinfo{person}{Ang Li}, {and} \bibinfo{person}{Yufei Ding}.} \bibinfo{year}{2021}\natexlab{}.
\newblock \showarticletitle{Apnn-tc: Accelerating arbitrary precision neural networks on ampere gpu tensor cores}. In \bibinfo{booktitle}{\emph{Proceedings of the international conference for high performance computing, networking, storage and analysis(SC)}}. \bibinfo{pages}{1--13}.
\newblock


\bibitem[Fey and Lenssen(2019)]%
        {fey2019fast}
\bibfield{author}{\bibinfo{person}{Matthias Fey} {and} \bibinfo{person}{Jan~Eric Lenssen}.} \bibinfo{year}{2019}\natexlab{}.
\newblock \showarticletitle{Fast graph representation learning with PyTorch Geometric}.
\newblock \bibinfo{journal}{\emph{arXiv preprint arXiv:1903.02428}} (\bibinfo{year}{2019}).
\newblock


\bibitem[Gale et~al\mbox{.}(2020)]%
        {sputnik}
\bibfield{author}{\bibinfo{person}{Trevor Gale}, \bibinfo{person}{Matei Zaharia}, \bibinfo{person}{Cliff Young}, {and} \bibinfo{person}{Erich Elsen}.} \bibinfo{year}{2020}\natexlab{}.
\newblock \showarticletitle{Sparse gpu kernels for deep learning}. In \bibinfo{booktitle}{\emph{International Conference for High Performance Computing, Networking, Storage and Analysis (SC)}}.
\newblock


\bibitem[Gibert et~al\mbox{.}(2012)]%
        {gibert2012graph}
\bibfield{author}{\bibinfo{person}{Jaume Gibert}, \bibinfo{person}{Ernest Valveny}, {and} \bibinfo{person}{Horst Bunke}.} \bibinfo{year}{2012}\natexlab{}.
\newblock \showarticletitle{Graph embedding in vector spaces by node attribute statistics}.
\newblock \bibinfo{journal}{\emph{Pattern Recognition}} \bibinfo{volume}{45}, \bibinfo{number}{9} (\bibinfo{year}{2012}), \bibinfo{pages}{3072--3083}.
\newblock


\bibitem[Hong et~al\mbox{.}(2019)]%
        {hong2019adaptive}
\bibfield{author}{\bibinfo{person}{Changwan Hong}, \bibinfo{person}{Aravind Sukumaran-Rajam}, \bibinfo{person}{Israt Nisa}, \bibinfo{person}{Kunal Singh}, {and} \bibinfo{person}{P Sadayappan}.} \bibinfo{year}{2019}\natexlab{}.
\newblock \showarticletitle{Adaptive sparse tiling for sparse matrix multiplication}. In \bibinfo{booktitle}{\emph{Proceedings of the 24th Symposium on Principles and Practice of Parallel Programming(PPoPP)}}. \bibinfo{pages}{300--314}.
\newblock


\bibitem[Huang et~al\mbox{.}(2020)]%
        {huang2020ge}
\bibfield{author}{\bibinfo{person}{Guyue Huang}, \bibinfo{person}{Guohao Dai}, \bibinfo{person}{Yu Wang}, {and} \bibinfo{person}{Huazhong Yang}.} \bibinfo{year}{2020}\natexlab{}.
\newblock \showarticletitle{$\{$GE-SpMM$\}$: General-purpose sparse matrix-matrix multiplication on gpus for graph neural networks}. In \bibinfo{booktitle}{\emph{International Conference for High Performance Computing, Networking, Storage and Analysis (SC)}}.
\newblock


\bibitem[Jiang et~al\mbox{.}(2020)]%
        {jiang2020novel}
\bibfield{author}{\bibinfo{person}{Peng Jiang}, \bibinfo{person}{Changwan Hong}, {and} \bibinfo{person}{Gagan Agrawal}.} \bibinfo{year}{2020}\natexlab{}.
\newblock \showarticletitle{A novel data transformation and execution strategy for accelerating sparse matrix multiplication on GPUs}. In \bibinfo{booktitle}{\emph{Proceedings of the 25th ACM SIGPLAN symposium on principles and practice of parallel programming(PPoPP)}}. \bibinfo{pages}{376--388}.
\newblock


\bibitem[Jiang and Luo(2022)]%
        {jiang2022graph}
\bibfield{author}{\bibinfo{person}{Weiwei Jiang} {and} \bibinfo{person}{Jiayun Luo}.} \bibinfo{year}{2022}\natexlab{}.
\newblock \showarticletitle{Graph neural network for traffic forecasting: A survey}.
\newblock \bibinfo{journal}{\emph{Expert Systems with Applications}}  \bibinfo{volume}{207} (\bibinfo{year}{2022}), \bibinfo{pages}{117921}.
\newblock


\bibitem[Khatua et~al\mbox{.}(2023)]%
        {khatua2023igb}
\bibfield{author}{\bibinfo{person}{Arpandeep Khatua}, \bibinfo{person}{Vikram~Sharma Mailthody}, \bibinfo{person}{Bhagyashree Taleka}, \bibinfo{person}{Tengfei Ma}, \bibinfo{person}{Xiang Song}, {and} \bibinfo{person}{Wen-mei Hwu}.} \bibinfo{year}{2023}\natexlab{}.
\newblock \showarticletitle{Igb: Addressing the gaps in labeling, features, heterogeneity, and size of public graph datasets for deep learning research}. In \bibinfo{booktitle}{\emph{Proceedings of the 29th ACM SIGKDD Conference on Knowledge Discovery and Data Mining(SIGKDD)}}. \bibinfo{pages}{4284--4295}.
\newblock


\bibitem[Khurana et~al\mbox{.}(2023)]%
        {khurana2023natural}
\bibfield{author}{\bibinfo{person}{Diksha Khurana}, \bibinfo{person}{Aditya Koli}, \bibinfo{person}{Kiran Khatter}, {and} \bibinfo{person}{Sukhdev Singh}.} \bibinfo{year}{2023}\natexlab{}.
\newblock \showarticletitle{Natural language processing: State of the art, current trends and challenges}.
\newblock \bibinfo{journal}{\emph{Multimedia tools and applications}} \bibinfo{volume}{82}, \bibinfo{number}{3} (\bibinfo{year}{2023}), \bibinfo{pages}{3713--3744}.
\newblock


\bibitem[Kipf and Welling(2016)]%
        {kipf2016semi}
\bibfield{author}{\bibinfo{person}{Thomas~N Kipf} {and} \bibinfo{person}{Max Welling}.} \bibinfo{year}{2016}\natexlab{}.
\newblock \showarticletitle{Semi-supervised classification with graph convolutional networks}.
\newblock \bibinfo{journal}{\emph{arXiv preprint arXiv:1609.02907}} (\bibinfo{year}{2016}).
\newblock


\bibitem[Kumar et~al\mbox{.}(2022)]%
        {kumar2022influence}
\bibfield{author}{\bibinfo{person}{Sanjay Kumar}, \bibinfo{person}{Abhishek Mallik}, \bibinfo{person}{Anavi Khetarpal}, {and} \bibinfo{person}{BS Panda}.} \bibinfo{year}{2022}\natexlab{}.
\newblock \showarticletitle{Influence maximization in social networks using graph embedding and graph neural network}.
\newblock \bibinfo{journal}{\emph{Information Sciences}}  \bibinfo{volume}{607} (\bibinfo{year}{2022}), \bibinfo{pages}{1617--1636}.
\newblock


\bibitem[Kunegis and Lommatzsch(2009)]%
        {kunegis2009learning}
\bibfield{author}{\bibinfo{person}{J{\'e}r{\^o}me Kunegis} {and} \bibinfo{person}{Andreas Lommatzsch}.} \bibinfo{year}{2009}\natexlab{}.
\newblock \showarticletitle{Learning spectral graph transformations for link prediction}. In \bibinfo{booktitle}{\emph{Proceedings of the 26th Annual International Conference on Machine Learning}}. \bibinfo{pages}{561--568}.
\newblock


\bibitem[Lan et~al\mbox{.}(2014)]%
        {lan2014sparse}
\bibfield{author}{\bibinfo{person}{Andrew~S Lan}, \bibinfo{person}{Andrew~E Waters}, \bibinfo{person}{Christoph Studer}, {and} \bibinfo{person}{Richard~G Baraniuk}.} \bibinfo{year}{2014}\natexlab{}.
\newblock \showarticletitle{Sparse factor analysis for learning and content analytics}.
\newblock \bibinfo{journal}{\emph{The Journal of Machine Learning Research}} \bibinfo{volume}{15}, \bibinfo{number}{1} (\bibinfo{year}{2014}), \bibinfo{pages}{1959--2008}.
\newblock


\bibitem[Li et~al\mbox{.}(2022)]%
        {magicube}
\bibfield{author}{\bibinfo{person}{Shigang Li}, \bibinfo{person}{Kazuki Osawa}, {and} \bibinfo{person}{Torsten Hoefler}.} \bibinfo{year}{2022}\natexlab{}.
\newblock \showarticletitle{Efficient quantized sparse matrix operations on tensor cores}. In \bibinfo{booktitle}{\emph{SC22: International Conference for High Performance Computing, Networking, Storage and Analysis (SC)}}. IEEE.
\newblock


\bibitem[Li(2022)]%
        {li2022research}
\bibfield{author}{\bibinfo{person}{Yinglong Li}.} \bibinfo{year}{2022}\natexlab{}.
\newblock \showarticletitle{Research and application of deep learning in image recognition}. In \bibinfo{booktitle}{\emph{2022 IEEE 2nd International Conference on Power, Electronics and Computer Applications (ICPECA)}}. IEEE, \bibinfo{pages}{994--999}.
\newblock


\bibitem[Lu and Liu(2023)]%
        {dasp}
\bibfield{author}{\bibinfo{person}{Yuechen Lu} {and} \bibinfo{person}{Weifeng Liu}.} \bibinfo{year}{2023}\natexlab{}.
\newblock \showarticletitle{DASP: Specific Dense Matrix Multiply-Accumulate Units Accelerated General Sparse Matrix-Vector Multiplication}. In \bibinfo{booktitle}{\emph{Proceedings of the International Conference for High Performance Computing, Networking, Storage and Analysis}}. \bibinfo{pages}{1--14}.
\newblock


\bibitem[Niu et~al\mbox{.}(2022)]%
        {niu2022tilespgemm}
\bibfield{author}{\bibinfo{person}{Yuyao Niu}, \bibinfo{person}{Zhengyang Lu}, \bibinfo{person}{Haonan Ji}, \bibinfo{person}{Shuhui Song}, \bibinfo{person}{Zhou Jin}, {and} \bibinfo{person}{Weifeng Liu}.} \bibinfo{year}{2022}\natexlab{}.
\newblock \showarticletitle{TileSpGEMM: A tiled algorithm for parallel sparse general matrix-matrix multiplication on GPUs}. In \bibinfo{booktitle}{\emph{Proceedings of the 27th ACM SIGPLAN Symposium on Principles and Practice of Parallel Programming(PPoPP)}}. \bibinfo{pages}{90--106}.
\newblock


\bibitem[{NVIDIA}({[n.\,d.]})]%
        {cuSPARSE}
\bibfield{author}{\bibinfo{person}{{NVIDIA}}.} \bibinfo{year}{[n.\,d.]}\natexlab{}.
\newblock \bibinfo{title}{{ cuSPARSE. }}.
\newblock
\newblock
\newblock
\shownote{https://developer.nvidia.cn/cusparse}.


\bibitem[{Nvidia}({[n.\,d.]})]%
        {mma}
\bibfield{author}{\bibinfo{person}{{Nvidia}}.} \bibinfo{year}{[n.\,d.]}\natexlab{}.
\newblock \bibinfo{title}{{Warp level matrix multiply-accumulate instructions. }}.
\newblock
\newblock
\newblock
\shownote{https://docs.nvidia.com/cuda/parallel-thread-execution/index.html\#matrix-shape}.


\bibitem[{NVIDIA}(2020)]%
        {cusparselt}
\bibfield{author}{\bibinfo{person}{{NVIDIA}}.} \bibinfo{year}{2020}\natexlab{}.
\newblock \bibinfo{title}{{Exploiting NVIDIA Ampere Structured Sparsity with cuSPARSELt }}.
\newblock
\newblock
\newblock
\shownote{https://developer.nvidia.com/blog/exploiting-ampere-structured-sparsity-with-cusparselt/}.


\bibitem[{NVIDIA}(2023)]%
        {m16n8k8}
\bibfield{author}{\bibinfo{person}{{NVIDIA}}.} \bibinfo{year}{2023}\natexlab{}.
\newblock \bibinfo{title}{{Matrix Fragments for mma.m16n8k8 }}.
\newblock
\newblock
\newblock
\shownote{https://docs.nvidia.com/cuda/parallel-thread-execution/index.html\#matrix-fragments-for-mma-m16n8k8}.


\bibitem[Pang et~al\mbox{.}(2024)]%
        {RoDe}
\bibfield{author}{\bibinfo{person}{Meng Pang}, \bibinfo{person}{Xiang Fei}, \bibinfo{person}{Peng Qu}, \bibinfo{person}{Youhui Zhang}, {and} \bibinfo{person}{Zhaolin Li}.} \bibinfo{year}{2024}\natexlab{}.
\newblock \showarticletitle{A Row Decomposition-based Approach for Sparse Matrix Multiplication on GPUs}. In \bibinfo{booktitle}{\emph{PPoPP24: Proceedings of the 29th ACM SIGPLAN Annual Symposium on Principles and Practice of Parallel Programming}}. \bibinfo{pages}{377--389}.
\newblock


\bibitem[Peng et~al\mbox{.}(2024)]%
        {peng2024maxk}
\bibfield{author}{\bibinfo{person}{Hongwu Peng}, \bibinfo{person}{Xi Xie}, \bibinfo{person}{Kaustubh Shivdikar}, \bibinfo{person}{Md~Amit Hasan}, \bibinfo{person}{Jiahui Zhao}, \bibinfo{person}{Shaoyi Huang}, \bibinfo{person}{Omer Khan}, \bibinfo{person}{David Kaeli}, {and} \bibinfo{person}{Caiwen Ding}.} \bibinfo{year}{2024}\natexlab{}.
\newblock \showarticletitle{Maxk-gnn: Extremely fast gpu kernel design for accelerating graph neural networks training}. In \bibinfo{booktitle}{\emph{Proceedings of the 29th ACM International Conference on Architectural Support for Programming Languages and Operating Systems, Volume 2}}. \bibinfo{pages}{683--698}.
\newblock


\bibitem[Piao et~al\mbox{.}(2022)]%
        {piao2022sparse}
\bibfield{author}{\bibinfo{person}{Yinhua Piao}, \bibinfo{person}{Sangseon Lee}, \bibinfo{person}{Dohoon Lee}, {and} \bibinfo{person}{Sun Kim}.} \bibinfo{year}{2022}\natexlab{}.
\newblock \showarticletitle{Sparse structure learning via graph neural networks for inductive document classification}. In \bibinfo{booktitle}{\emph{Proceedings of the AAAI Conference on Artificial Intelligence(AAAI)}}, Vol.~\bibinfo{volume}{36}. \bibinfo{pages}{11165--11173}.
\newblock


\bibitem[Pool et~al\mbox{.}(2021)]%
        {SparseTC}
\bibfield{author}{\bibinfo{person}{Jeff Pool}, \bibinfo{person}{Abhishek Sawarkar}, {and} \bibinfo{person}{Jay Rodge}.} \bibinfo{year}{2021}\natexlab{}.
\newblock \bibinfo{title}{Accelerating Inference with Sparsity Using the NVIDIA Ampere Architecture and NVIDIA TensorRT}.
\newblock
\newblock


\bibitem[Sala et~al\mbox{.}(2010)]%
        {sala2010brief}
\bibfield{author}{\bibinfo{person}{Alessandra Sala}, \bibinfo{person}{Haitao Zheng}, \bibinfo{person}{Ben~Y Zhao}, \bibinfo{person}{Sabrina Gaito}, {and} \bibinfo{person}{Gian~Paolo Rossi}.} \bibinfo{year}{2010}\natexlab{}.
\newblock \showarticletitle{Brief announcement: revisiting the power-law degree distribution for social graph analysis}. In \bibinfo{booktitle}{\emph{Proceedings of the 29th ACM SIGACT-SIGOPS symposium on Principles of distributed computing}}. \bibinfo{pages}{400--401}.
\newblock


\bibitem[Scarselli et~al\mbox{.}(2008)]%
        {scarselli2008graph}
\bibfield{author}{\bibinfo{person}{Franco Scarselli}, \bibinfo{person}{Marco Gori}, \bibinfo{person}{Ah~Chung Tsoi}, \bibinfo{person}{Markus Hagenbuchner}, {and} \bibinfo{person}{Gabriele Monfardini}.} \bibinfo{year}{2008}\natexlab{}.
\newblock \showarticletitle{The graph neural network model}.
\newblock \bibinfo{journal}{\emph{IEEE transactions on neural networks}} \bibinfo{volume}{20}, \bibinfo{number}{1} (\bibinfo{year}{2008}), \bibinfo{pages}{61--80}.
\newblock


\bibitem[Thekumparampil et~al\mbox{.}(2018)]%
        {thekumparampil2018attention}
\bibfield{author}{\bibinfo{person}{Kiran~K Thekumparampil}, \bibinfo{person}{Chong Wang}, \bibinfo{person}{Sewoong Oh}, {and} \bibinfo{person}{Li-Jia Li}.} \bibinfo{year}{2018}\natexlab{}.
\newblock \showarticletitle{Attention-based graph neural network for semi-supervised learning}.
\newblock \bibinfo{journal}{\emph{arXiv preprint arXiv:1803.03735}} (\bibinfo{year}{2018}).
\newblock


\bibitem[Tylenda et~al\mbox{.}(2009)]%
        {tylenda2009towards}
\bibfield{author}{\bibinfo{person}{Tomasz Tylenda}, \bibinfo{person}{Ralitsa Angelova}, {and} \bibinfo{person}{Srikanta Bedathur}.} \bibinfo{year}{2009}\natexlab{}.
\newblock \showarticletitle{Towards time-aware link prediction in evolving social networks}. In \bibinfo{booktitle}{\emph{Proceedings of the 3rd workshop on social network mining and analysis}}. \bibinfo{pages}{1--10}.
\newblock


\bibitem[Veli{\v{c}}kovi{\'c} et~al\mbox{.}(2017)]%
        {velivckovic2017graph}
\bibfield{author}{\bibinfo{person}{Petar Veli{\v{c}}kovi{\'c}}, \bibinfo{person}{Guillem Cucurull}, \bibinfo{person}{Arantxa Casanova}, \bibinfo{person}{Adriana Romero}, \bibinfo{person}{Pietro Lio}, {and} \bibinfo{person}{Yoshua Bengio}.} \bibinfo{year}{2017}\natexlab{}.
\newblock \showarticletitle{Graph attention networks}.
\newblock \bibinfo{journal}{\emph{arXiv preprint arXiv:1710.10903}} (\bibinfo{year}{2017}).
\newblock


\bibitem[Wang et~al\mbox{.}(2019)]%
        {wang2019deep}
\bibfield{author}{\bibinfo{person}{Minjie Wang}, \bibinfo{person}{Da Zheng}, \bibinfo{person}{Zihao Ye}, \bibinfo{person}{Quan Gan}, \bibinfo{person}{Mufei Li}, \bibinfo{person}{Xiang Song}, \bibinfo{person}{Jinjing Zhou}, \bibinfo{person}{Chao Ma}, \bibinfo{person}{Lingfan Yu}, \bibinfo{person}{Yu Gai}, {et~al\mbox{.}}} \bibinfo{year}{2019}\natexlab{}.
\newblock \showarticletitle{Deep graph library: A graph-centric, highly-performant package for graph neural networks}.
\newblock \bibinfo{journal}{\emph{arXiv preprint arXiv:1909.01315}} (\bibinfo{year}{2019}).
\newblock


\bibitem[Wang et~al\mbox{.}(2021)]%
        {wang2021gnnadvisor}
\bibfield{author}{\bibinfo{person}{Yuke Wang}, \bibinfo{person}{Boyuan Feng}, \bibinfo{person}{Gushu Li}, \bibinfo{person}{Shuangchen Li}, \bibinfo{person}{Lei Deng}, \bibinfo{person}{Yuan Xie}, {and} \bibinfo{person}{Yufei Ding}.} \bibinfo{year}{2021}\natexlab{}.
\newblock \showarticletitle{$\{$GNNAdvisor$\}$: An adaptive and efficient runtime system for $\{$GNN$\}$ acceleration on $\{$GPUs$\}$}. In \bibinfo{booktitle}{\emph{15th USENIX symposium on operating systems design and implementation (OSDI)}}.
\newblock


\bibitem[Wang et~al\mbox{.}(2023)]%
        {wang2023tc}
\bibfield{author}{\bibinfo{person}{Yuke Wang}, \bibinfo{person}{Boyuan Feng}, \bibinfo{person}{Zheng Wang}, \bibinfo{person}{Guyue Huang}, {and} \bibinfo{person}{Yufei Ding}.} \bibinfo{year}{2023}\natexlab{}.
\newblock \showarticletitle{$\{$TC-GNN$\}$: Bridging Sparse $\{$GNN$\}$ Computation and Dense Tensor Cores on $\{$GPUs$\}$}. In \bibinfo{booktitle}{\emph{2023 USENIX Annual Technical Conference (USENIX ATC)}}.
\newblock


\bibitem[Wu et~al\mbox{.}(2023)]%
        {wu2023turbomgnn}
\bibfield{author}{\bibinfo{person}{Wenchao Wu}, \bibinfo{person}{Xuanhua Shi}, \bibinfo{person}{Ligang He}, {and} \bibinfo{person}{Hai Jin}.} \bibinfo{year}{2023}\natexlab{}.
\newblock \showarticletitle{TurboMGNN: Improving Concurrent GNN Training Tasks on GPU With Fine-Grained Kernel Fusion}.
\newblock \bibinfo{journal}{\emph{IEEE Transactions on Parallel and Distributed Systems (TPDS)}} (\bibinfo{year}{2023}).
\newblock


\bibitem[Zachariadis et~al\mbox{.}(2020)]%
        {zachariadis2020accelerating}
\bibfield{author}{\bibinfo{person}{Orestis Zachariadis}, \bibinfo{person}{Nitin Satpute}, \bibinfo{person}{Juan G{\'o}mez-Luna}, {and} \bibinfo{person}{Joaqu{\'\i}n Olivares}.} \bibinfo{year}{2020}\natexlab{}.
\newblock \showarticletitle{Accelerating sparse matrix--matrix multiplication with GPU Tensor Cores}.
\newblock \bibinfo{journal}{\emph{Computers \& Electrical Engineering}}  \bibinfo{volume}{88} (\bibinfo{year}{2020}), \bibinfo{pages}{106848}.
\newblock


\bibitem[Zeng et~al\mbox{.}(2019)]%
        {zeng2019graphsaint}
\bibfield{author}{\bibinfo{person}{Hanqing Zeng}, \bibinfo{person}{Hongkuan Zhou}, \bibinfo{person}{Ajitesh Srivastava}, \bibinfo{person}{Rajgopal Kannan}, {and} \bibinfo{person}{Viktor Prasanna}.} \bibinfo{year}{2019}\natexlab{}.
\newblock \showarticletitle{Graphsaint: Graph sampling based inductive learning method}.
\newblock \bibinfo{journal}{\emph{arXiv preprint arXiv:1907.04931}} (\bibinfo{year}{2019}).
\newblock


\end{thebibliography}

\end{document}